\documentclass[final,3p,times]{elsarticle}

\journal{Parallel Computing}
\usepackage{marvosym}
\usepackage{amsmath,amssymb,amsfonts}
\usepackage{stmaryrd}

\usepackage[english]{babel}
\usepackage[utf8]{inputenc}
\usepackage{hyperref}
\usepackage{graphicx}

\usepackage[usenames,dvipsnames]{color}
\usepackage{tikz}
\usepackage{pgfplotstable}

\usepackage{placeins}
\usepackage{float}
\usepackage{caption}

\usepackage{siunitx}

\usepackage{array,multirow}

\usepackage{empheq} % for aligned with subequations

\usetikzlibrary{plotmarks,backgrounds,shadows,spy,fixedpointarithmetic,patterns,intersections,arrows.meta,shapes,decorations.text,decorations.pathmorphing,backgrounds,fit,positioning,shapes.symbols,chains,3d,calc,external}
\usepackage{fp}

\newcommand\vE{\mathbf{E}}
\newcommand\vF{\mathbf{F}}

\newcommand\eps{\varepsilon}
\newcommand{\vn} {{\bf n}}
\newcommand{\vv} {{\bf v}}

\definecolor{corrections}{rgb}{0,0,0}

\makeatletter
\def\ps@pprintTitle{%
   \let\@oddhead\@empty
   \let\@evenhead\@empty
   \let\@oddfoot\@empty
   \let\@evenfoot\@oddfoot
}
\makeatother

\begin{document}

\begin{frontmatter}



\title{Microwave Tomographic Imaging of Cerebrovascular Accidents by Using High-Performance Computing}

\author[label1]{P.-H.~Tournier}
\author[label3]{I.~Aliferis}
\author[label10]{M.~Bonazzoli}
\author[label5]{M.~de~Buhan}
\author[label7]{M.~Darbas}
\author[label2,label4]{V.~Dolean}
\author[label1]{F.~Hecht}
\author[label6]{P.~Jolivet}
\author[label3]{I.~El~Kanfoud}
\author[label3]{C.~Migliaccio}
\author[label1]{F.~Nataf}
\author[label3,label9]{Ch.~Pichot}
\author[label8]{S.~Semenov}

\address[label1]{Sorbonne Universit\'e, Universit\'e Paris-Diderot SPC, CNRS, Inria, Laboratoire Jacques-Louis Lions, \'equipe Alpines, F-75005 Paris}
%\address[label1a]{Laboratoire Jacques-Louis Lions, UMR CNRS 7598, Sorbonne Universit\'es, UPMC, Paris, France}
%\address[label1b]{INRIA-Paris, EPC Alpines, Paris, France}
\address[label3]{Universit\'e C\^ote d'Azur, CNRS, LEAT, France}
\address[label2]{Universit\'e C\^ote d'Azur, CNRS, LJAD, France}
\address[label5]{MAP5, UMR CNRS 8145, Universit\'e Paris-Descartes, Sorbonne Paris Cit\'e, France}
\address[label7]{LAMFA, UMR CNRS 7352, Universit\'e de Picardie Jules Verne, Amiens, France}
\address[label4]{Dept of Maths and Stats, University of Strathclyde, Glasgow, UK}
\address[label6]{IRIT, UMR CNRS 5505, Toulouse, France}
\address[label8]{EMTensor GmbH, TechGate, 1220 Vienna, Austria}
\address[label9]{School of Innovation, Design and Engineering, M{\"a}lardalen University, Sweden}
\address[label10]{INRIA Saclay Île-de-France, CMAP, École Polytechnique, Palaiseau, France}

%\author{}

%\address{}

\begin{abstract}
The motivation of this work is the detection of cerebrovascular accidents by microwave tomographic imaging. This requires the solution of an inverse problem relying on a minimization algorithm (for example, gradient-based), where successive iterations consist in repeated solutions of a direct problem. The reconstruction algorithm is extremely computationally intensive and makes use of efficient parallel algorithms and high-performance computing. The feasibility of this type of imaging is conditioned on one hand by an accurate reconstruction of the material properties of the propagation medium and on the other hand by a considerable reduction in simulation time. Fulfilling these two requirements will enable a very rapid and accurate diagnosis. From the mathematical and numerical point of view, this means solving Maxwell's equations in time-harmonic regime by appropriate domain decomposition methods, which are naturally adapted to parallel architectures.
\end{abstract}


\end{frontmatter}


\section{Introduction}
\label{sec:introduction}

A stroke, also known as cerebrovascular accident, is a disturbance in the blood supply to the brain caused by a blocked or burst blood vessel. As a consequence, cerebral tissues are deprived of oxygen and nutrients. This results in a rapid loss of brain functions and often death. Strokes are classified into two major categories: ischemic ($85$\% of strokes) and hemorrhagic ($15$\% of strokes). During an acute ischemic stroke, the blood supply to a part of the brain is interrupted by thrombosis - the formation of a blood clot in a blood vessel - or by an embolism elsewhere in the body. A hemorrhagic stroke occurs when a blood vessel bursts inside the brain, increasing pressure in the brain and injuring brain cells. The two types of strokes result in opposite variations of the dielectric properties of the affected tissues. How quickly one can detect and characterize the stroke is of fundamental importance for the survival of the patient. The quicker the treatment is, the more reversible the damage and the better the chances of recovery are. Moreover, the treatment of ischemic stroke consists in thinning the blood (anticoagulants) and can be fatal if the stroke is hemorrhagic. 
%Conversely, giving the treatment for a hemorrhagic stroke to a patient with ischemic stroke can kill the patient. 
Therefore, it is vital to make a clear distinction between the two types of strokes before treating the patient. Moreover, ideally one would want to monitor continuously the effect of the treatment on the evolution of the stroke during the hospitalization. The two most used imaging techniques for strokes diagnosis are MRI (magnetic resonance imaging) and CT scan (computerized tomography scan). One of their downsides is that the travel time from the patient's home to the hospital is lost. Moreover, the cost and the lack of portability of MRI and the harmful character of CT scan\textcolor{corrections}{, which uses ionizing radiation and thus cannot be used repeatedly,} make them unsuitable for a continuous monitoring at the hospital during treatment.

This has motivated the study of an additional technique: microwave tomography. The measurement system is lightweight and thus transportable. The acquisition of the data is harmless and faster than CT or MRI. \textcolor{corrections}{Microwave technology offers the potential for a low-cost, non-invasive modality in a non-ionizing range of the frequency spectrum. Microwave imaging uses low power microwave signals of the order of $1$\,mWatt transmitted towards the head during $2.0$ to $2.5$ seconds, orders of magnitude less than the power by a cell phone during a phone call.} Hence, this imaging modality could be used by an emergency unit and for monitoring at the hospital. At frequencies of the order of \SI{1}{\giga\hertz}, the tissues are well differentiated and can be imaged on the basis of their dielectric properties. After the first works on microwave imaging in 1982 by Lin and Clarke \cite{lin1982microwave}, other works followed, but almost always on synthetic simplified models \cite{Semenov:2008:MTB}. New devices are currently designed and studied %by the University of Chalmers (Gothenburg, Sweden) \cite{Persson:2014:MBS} and 
 by EMTensor GmbH (Vienna, Austria) \cite{Semenov:2014:ETB}.

\textcolor{corrections}{There are mainly two classes of microwave imaging methods: qualitative and quantitative imaging algorithms. The first one is a linearized inversion synthetic-aperture-radar method based on delay-and-sum \cite{mustafa2013novel}, beamforming \cite{li2005overview} or backpropagation or backprojection algorithm \cite{zamani2016fast}. The second one is a nonlinear method relying on the minimization of a cost functional, which depends on the discrepancy between the experimental data and data simulated by a forward time-harmonic Maxwell's model \cite{pichot1997gradient,irishina2013brain}. The minimization of the functional is usually carried out by a Newton-type algorithm or gradient method \cite{pichot1997gradient,irishina2013brain}. However, fast diagnosis is essential to save the patient. This is why fast inversion techniques based on qualitative methods have been developed, but they cannot differentiate between ischemic and hemorrhagic strokes, which correspond to an opposite variation in dielectric contrast. Therefore, there is a strong need for fast quantitative imaging techniques, allowing a rapid diagnosis and identification of the type of stroke as well as monitoring during the treatment.}

\begin{figure}
\centering
\includegraphics[height=0.17\textheight]{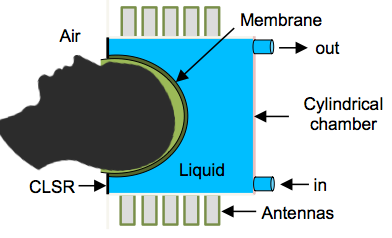} 
\includegraphics[height=0.17\textheight]{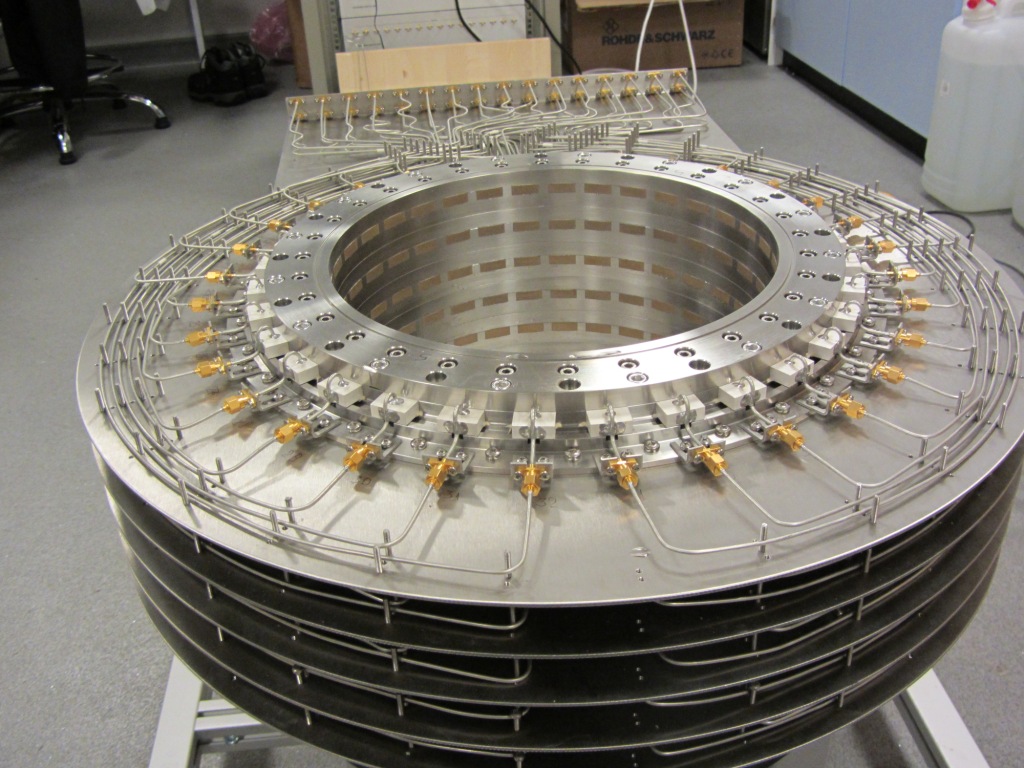} \quad
\includegraphics[height=0.17\textheight]{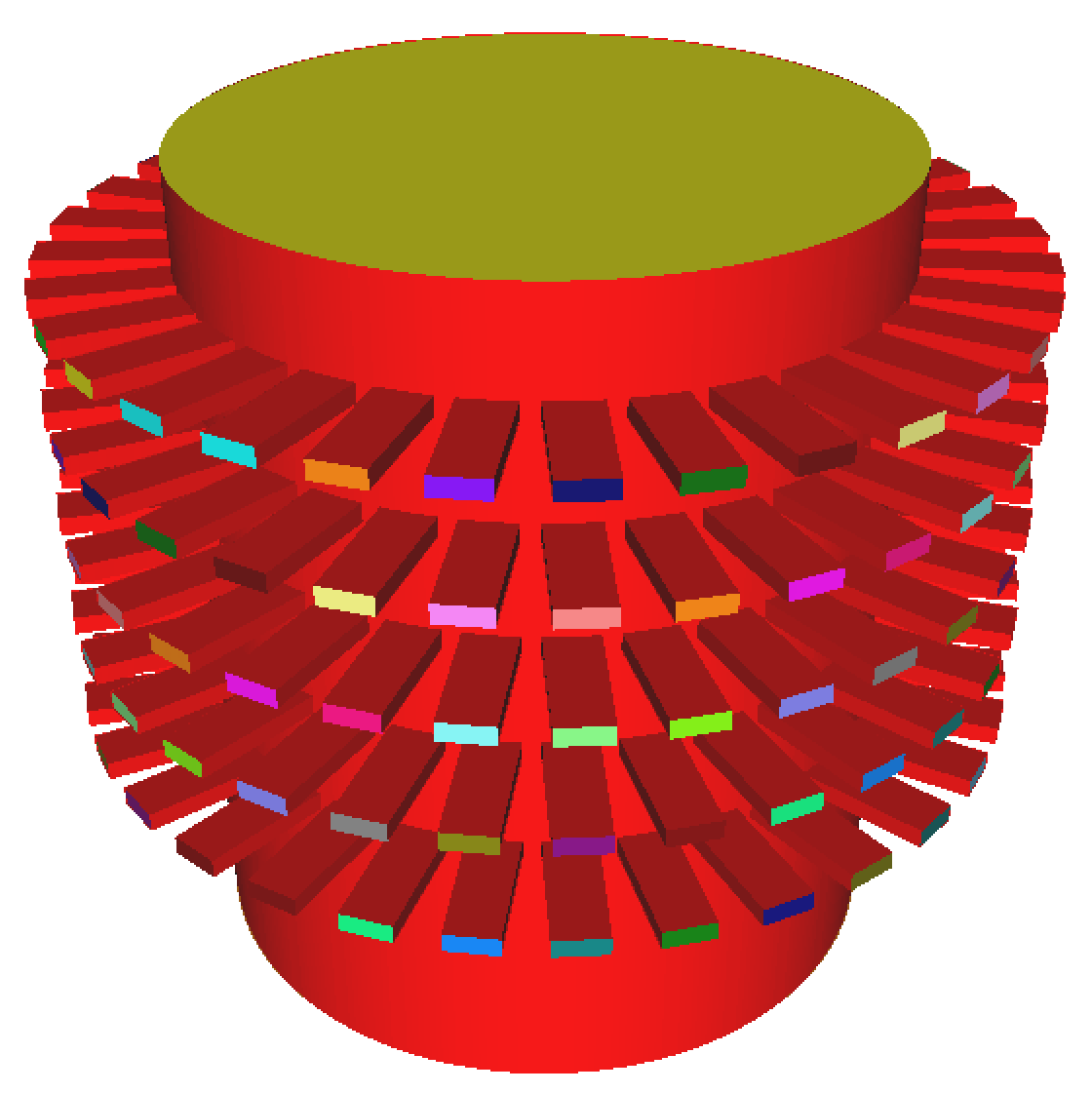} 
\caption{Left: Operating principle of the diagnosis apparatus. Middle: imaging system prototype of EMTensor, by courtesy of EMTensor company. Right: the corresponding simulation domain.} 
\label{fig:chamber}
\end{figure}

\begin{figure}
\centering 
\includegraphics[width=0.7\textwidth]{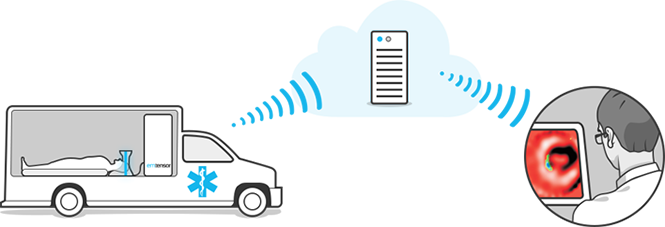}
%\caption{Principe de la technologie de diagnostique.} 
\caption{Design concept of the diagnosis technology, by courtesy of EMTensor company.}
\label{fig:tecno}
\end{figure}

The purpose of this work is to solve in parallel the \textcolor{corrections}{nonlinear} inverse problem associated with the time-harmonic Maxwell's equations, which model electromagnetic waves propagation. The dielectric properties of the brain tissues of a patient yield the image that could be used for a rapid diagnosis of brain strokes. Simulation results presented in this work have been obtained on the imaging system prototype developed by EMTensor GmbH \cite{Semenov:2014:ETB} (see Figure~\ref{fig:chamber}). It is composed of $5$ rings of $32$ \textcolor{corrections}{antennas, which are} ceramic-loaded rectangular waveguides, around a metallic cylindrical chamber of diameter $28.5$\,cm and total height $28$\,cm. The head of the patient is inserted into the chamber as shown in Figure~\ref{fig:chamber} (left). The imaging chamber is filled with a matching solution and a membrane is used to isolate the head. 
Each antenna successively transmits a signal at a fixed frequency, typically \SI{1}{\giga\hertz}.
%The working frequency is typically \SI{1}{\giga\hertz}. 
%Each antenna can act as transmitter and receiver. 
The electromagnetic wave propagates inside the chamber and in the object to be imaged according to its electromagnetic properties. The retrieved data then consist in the scattering parameters measured by the $160$ receiving antennas, which are used as input for the inverse problem. \textcolor{corrections}{The amount of input data ($160\times160$ complex numbers) is minimal and} can be wirelessly transferred 
%via a rapid telephony standard such as 4G and 5G 
to a remote computing center. The HPC machine will then compute the 3D images of the patient's brain. 
%It is worth noting that the supercomputer need not be run by the hospital and could be used for other computations as well. 
Once formed, these images can be quickly transmitted from the computing center to the hospital, see Figure~\ref{fig:tecno}.


The paper is organized as follows. 
\textcolor{corrections}{In Section~\ref{sec:direct} the direct (or forward) problem, given by the time-harmonic Maxwell's equations in curl-curl form with suitable boundary conditions, is introduced and its finite element discretization is briefly described.} 
%In Section~\ref{sec:direct} the direct (or forward) problem and the time-harmonic Maxwell's equations in curl-curl form with suitable boundary conditions are introduced. 
%In Section~\ref{sec:edgelements}, we briefly describe the discretization method with edge finite elements. 
Section~\ref{sec:domaindecomposition} is devoted to the \textcolor{corrections}{construction} of the domain decomposition preconditioner \textcolor{corrections}{for the linear systems resulting from the discretization}. In Section~\ref{sec:scattering} we explain how to compute the scattering parameters. We also compare measurement data obtained by EMTensor with the coefficients computed by the simulation. We \textcolor{corrections}{derive and discuss} the inverse problem in Section~\ref{sec:inverseproblem}. Section~\ref{sec:numericalresults} is dedicated to numerical results. We first perform a strong scaling analysis to show the effectiveness of the domain decomposition method. Then, we present results obtained by solving the inverse problem in a realistic configuration, with noisy synthetic data generated using a numerical brain model with a simulated hemorrhagic stroke. Finally, we conclude this paper in Section~\ref{sec:conclusion} and give directions for future research.

\section{The direct problem}
\label{sec:direct}
Let the domain $\Omega \subset \mathbb{R}^3$ represent the imaging chamber (see Figure~\ref{fig:chamber}, right). We consider in $\Omega$ a heterogeneous non-magnetic dissipative linear isotropic dielectric medium, of dielectric permittivity $\eps({\bf x}) > 0$ and electrical conductivity $\sigma({\bf x}) \ge 0$. For each transmitting antenna $j = 1,\dots,N$ emitting a time periodic signal at angular frequency $\omega$, the complex amplitude $\vE_j({\bf x})$ of the associated electric field ${\bf \cal E}_j({\bf x},{\bf t}) = \Re (\vE_j({\bf x}) e^{{\text i}\omega {\bf t}})$ is solution to the following second order time-harmonic Maxwell's equation:
\begin{equation}
\label{eqdomain}
\nabla \times (\nabla \times \vE_j) -  \mu_0 (\omega^2 {\eps} - \text{i} \omega {\sigma}) \vE_j = {0}  \quad\mbox{ in } \Omega,
\end{equation}
where $\mu_0$ is the permeability of free space. Note that the coefficient $\kappa = \mu_0 (\omega^2 \eps - \text{i} \omega \sigma)$ in the equation can be written as $\kappa = \omega^2 \mu_0 \bigl(\eps - \text{i} \frac{\sigma}{\omega}\bigr)$, and in the next sections we will consider the relative complex permittivity $\eps_r$ given by the relation $\eps_r \eps_0 = \eps - \text{i} \frac{\sigma}{\omega}$, where $\eps_0$ is the permittivity of free space. Let $\vn$ be the unit outward normal to $\partial \Omega$. 
%Equation (\ref{eqdomain}) is equipped with perfectly conducting boundary conditions on the metallic walls $\Gamma_\text{m}$:
%\begin{equation*}
%% \label{eqmetal}
%\vE_j \times \vn= {0} \quad\mbox{ on } {\color{black}\Gamma_\text{m}},
%\end{equation*}
%and with impedance boundary conditions on the outer section of the transmitting waveguide $j$ and of the receiving waveguides $i = 1,\dots,N$, $i \neq j$ (see e.g.~\cite{beck1999multilevel}):
%%\begin{equation}
%\begin{alignat}{2}
%\label{eqwgj}
%(\nabla \times \vE_j) \times \vn +\text{i} \beta \vn \times (\vE_j \times \vn) &= {{\bf{g}}_j} &&\quad\mbox{ on }  {\color{black}\Gamma_j}, \\
%\label{eqwgi}
%(\nabla \times \vE_j) \times \vn +\text{i} \beta \vn \times (\vE_j \times \vn) &= {0} &&\quad\mbox{ on } {\color{black}\Gamma_i} \mbox{ , } i \neq j.
%\end{alignat}
%%\end{equation}
\textcolor{corrections}{Equation (\ref{eqdomain}) is equipped with perfectly conducting boundary conditions~\eqref{eqmetal} on the metallic walls $\Gamma_\text{m}$, 
and with impedance boundary conditions~\eqref{eqwgj}--\eqref{eqwgi} on the outer section $\Gamma_j$, resp.~$\Gamma_i$, of the transmitting waveguide $j$, resp.~receiving waveguides $i = 1,\dots,N$, $i \neq j$ (see e.g.~\cite{beck1999multilevel}):}
\begin{subequations}
\label{pbdirect}
\begin{empheq}[left=\empheqlbrace]{align}
\nabla \times (\nabla \times \vE_j) -  \mu_0 (\omega^2 {\eps} - \text{i} \omega {\sigma}) \vE_j &= {0} \quad   \mbox{ in } \Omega, \\
\vE_j \times \vn&= {0} \quad \mbox{ on } {\Gamma_\text{m}},  \label{eqmetal}\\
(\nabla \times \vE_j) \times \vn +\text{i} \beta \vn \times (\vE_j \times \vn) &= {{\bf{g}}_j} \quad \mbox{ on }  {\Gamma_j}, \label{eqwgj}\\
(\nabla \times \vE_j) \times \vn +\text{i} \beta \vn \times (\vE_j \times \vn) &= {0} \quad \mbox{ on } {\Gamma_i} \mbox{ , } i \neq j. \label{eqwgi}
\end{empheq} 
\end{subequations}
Here $\beta$ is the propagation wavenumber along the waveguide, corresponding to the propagation of the TE$_{10}$ fundamental mode. Equation (\ref{eqwgj}) imposes an incident wave which corresponds to the excitation of the fundamental mode $\vE_j^0$ of the $j$-th waveguide, with ${\bf{g}}_j = (\nabla \times \vE_j^0) \times \vn +\text{i} \beta  \vn \times (\vE_j^0 \times \vn)$. On the other hand equation (\ref{eqwgi}) corresponds to a first order absorbing boundary condition of Silver--M\"uller approximating a transparent boundary condition on the outer section of the receiving waveguides $i = 1,\dots,N$, $i \neq j$.
The bottom of the chamber is metallic, and we impose an impedance boundary condition on the top of the chamber. 
%We end up with the following boundary value problem for each transmitting antenna $j = 1,\dots,N$: find $\vE_j$ such that
%\begin{equation}
%\quad \left\{
%\label{pbdirect}
%\begin{aligned}
%\nabla \times (\nabla \times \vE_j) -  \mu_0 (\omega^2 {\color{black}\eps} - \text{i} \omega {\color{black}\sigma}) \vE_j &= {0}   &&\mbox{ in } \Omega, \\
%\vE_j \times \vn&= {0}  &&\mbox{ on } {\color{black}\Gamma_\text{m}}, \\
%(\nabla \times \vE_j) \times \vn +\text{i} \beta \vn \times (\vE_j \times \vn) &= {{\bf{g}}_j} &&\mbox{ on }  {\color{black}\Gamma_j}, \\
%(\nabla \times \vE_j) \times \vn +\text{i} \beta \vn \times (\vE_j \times \vn) &= {0} &&\mbox{ on } {\color{black}\Gamma_i} \mbox{ , } i \neq j.
%\end{aligned}
%\right.
%\end{equation}

Now, let $V = \{\vv \in H(\text{curl},\Omega), \vv \times \vn = 0 \text{ on } \Gamma_\text{m}, \vv \times \vn \in L_2(\cup_{i=1}^N \Gamma_i)^3\}$, where $H(\text{curl},\Omega) = \{\mathbf{v} \in L_2(\Omega)^3, \nabla \times \mathbf{v} \in L_2(\Omega)^3 \}$ is the space of square integrable functions whose curl is also square integrable. For each transmitting antenna $j = 1,\dots,N$, \textcolor{corrections}{the variational form of the associated direct (or forward) problem}~\eqref{pbdirect} reads: find $\vE_j \in V$ such that
\begin{equation}
\label{pbdirectvarf}
\begin{aligned}
\int_{\Omega} \left[ (\nabla \times \vE_j) \cdot (\nabla \times \vv) - \mu_0 (\omega^2 {\color{black}\eps} - \text{i} \omega {\color{black}\sigma}) \vE_j \cdot \vv \right]
 + \int_{\bigcup_{i=1}^N \Gamma_i} \text{i} \beta (\vE_j \times \vn) \cdot  (\vv \times \vn)
= \int_{\Gamma_j} {\bf{g}}_j \cdot \vv \quad \forall \vv \in V.
\end{aligned}
\end{equation}

\subsection{Edge finite element discretization}
\label{subsec:edgelements}

\textcolor{corrections}{In order to discretize problem~\eqref{pbdirectvarf} by a finite element method, consider a tetrahedral mesh $\mathcal{T}$ of the computational domain $\Omega$. 
N\'ed\'elec edge elements \cite{Nedelec:1980:MFE} are finite elements particularly suited for the approximation of the electric field. 
Indeed, they ensure continuity of the tangential component of the field and the finite dimensional subspace $V_h$ generated by N\'ed\'elec basis functions is included in $H(\text{curl},\Omega)$.  
N\'ed\'elec elements are called edge elements because, at the lowest order (degree $r=1$), basis functions and degrees of freedom are associated with the (oriented) edges of the mesh $\mathcal{T}$: the degrees of freedom are circulations of the field along the edges.} 

\textcolor{corrections}{The finite element discretization of the variational problem~\eqref{pbdirectvarf} produces linear systems 
\begin{equation}
\label{eq:linsys}
A \textbf{u}_j = \textbf{b}_j, 
\end{equation}
one for each transmitting antenna $j = 1,\dots,N$. Note that the matrix $A$ is the same for all transmitting antennas, but the right-hand side $\textbf{b}_j$ is different. }

\section{Domain decomposition preconditioning}
\label{sec:domaindecomposition}
%The finite element discretization of the variational problem~\eqref{pbdirectvarf} produces linear systems 
%\begin{equation*}% \label{eq:linsys}
%A {\bf u}_j = {\bf b}_j
%\end{equation*}
%for each transmitting antenna $j$. 
\textcolor{corrections}{Since the matrix $A$ of the linear systems~\eqref{eq:linsys} can be ill-conditioned, we need a robust and efficient preconditioner for the iterative solver (GMRES).} 
%This, combined with the fact that the underlying PDE is indefinite, highlights the need for a robust and efficient preconditioner. 
Here we employ domain decomposition preconditioners, which are extensively described in \cite{Dolean:2015:IDD}, as they are naturally suited to parallel computing. \textcolor{corrections}{The construction of the chosen} domain decomposition preconditioner is presented in the following.

%Let $\mathcal{T}$ be the mesh of the computational domain $\Omega$. 

\textcolor{corrections}{First, the mesh $\mathcal{T}$ is partitioned into $N_S$ non-overlapping meshes $\{\mathcal{T}_i\}_{1 \leqslant i \leqslant N_S}$ using standard graph partitioners such as SCOTCH~\cite{pellegrini1996scotch} or METIS~\cite{karypis1998fast}. If $\delta$ is a positive integer, the \emph{overlapping} decomposition $\{\mathcal{T}^\delta_i\}_{1 \leqslant i \leqslant N_S}$ is defined recursively as follows: $\mathcal{T}^\delta_i$ is obtained by including all tetrahedra of $\mathcal{T}^{\delta - 1}_i$ plus all adjacent tetrahedra of~$\mathcal{T}^{\delta - 1}_i$; for $\delta = 0$, $\mathcal{T}^{\delta}_i = \mathcal{T}_i$. 
Note that the number of layers in the overlap is then~$2\delta$.
%Let $V_h$ be the edge finite element space defined on $\mathcal{T}$, and $\{V^\delta_i\}_{1 \leqslant i \leqslant N_S}$ the local edge finite element spaces defined on $\{\mathcal{T}^\delta_i\}_{1 \leqslant i \leqslant N_S}$, $\delta>0$. 
Now, let $\{V^\delta_i\}_{1 \leqslant i \leqslant N_S}$  be the local edge finite element spaces defined on $\{\mathcal{T}^\delta_i\}_{1 \leqslant i \leqslant N_S}$, $\delta>0$.  
Consider the restrictions $\{R_i\}_{1 \leqslant i \leqslant N_S}$ from $V_h$ to $\{V^\delta_i\}_{1 \leqslant i \leqslant N_S}$, and a local partition of unity $\{D_i\}_{1 \leqslant i \leqslant N_S}$ such that}
\begin{equation}
\label{eq:algebPartUnity}
\sum_{i = 1}^{N_S} R_i^T D_i R_i = I_{n\times n}.
\end{equation}
Algebraically speaking, if $n$ is the global number of unknowns and $\{n_i\}_{1\leqslant i \leqslant N_S}$ are the numbers of unknowns for each local finite element space, then $R_i$ is a Boolean matrix of size $n_i \times n$, and $D_i$ is a diagonal matrix of size $n_i \times n_i$, for all $1\leqslant i \leqslant N_S$.
Note that $R_i^T$,  the transpose of $R_i$, is a $n \times n_i$ matrix that gives the extension by $0$ from $V^\delta_i$ to $V_h$.

Using these matrices, one can define the following one-level preconditioner, called Optimized Restricted Additive Schwarz preconditioner (ORAS) \cite{StCyr:2007:OMA,DJR:1992:ddmhm}: 
\begin{equation}
\label{eq:ORAS}
{M}^{-1}_{\text{ORAS}} = \sum_{i = 1}^{N_S} R_i^T D_i B_i^{-1} R_i,
\end{equation}
\textcolor{corrections}{where $\{B_i\}_{1 \leqslant i \leqslant N_S}$ are local operators corresponding to the subproblems with impedance boundary conditions $(\nabla\times\mathbf{E})\times\mathbf{n} + {\tt i} k \mathbf{n} \times 
(\mathbf{E} \times \mathbf{n})$ on the interfaces between subdomains, where $k = \omega\sqrt{\mu_0 \varepsilon}$ is the wavenumber.}
%These boundary conditions were first used as transmission conditions at the interfaces between subdomains in~\cite{DJR:1992:ddmhm}.
%The local matrices $\{B_i\}_{1 \leqslant i \leqslant N_S}$ of the ORAS preconditioner make use of more efficient transmission boundary conditions than the submatrices $\{R_i A R_i^T\}_{1 \leqslant i \leqslant N_S}$ of the original Restricted Additive Schwarz (RAS) preconditioner \cite{CaiSar:1999:RAS}.
It is important to note that when a direct solver is used to compute the action of $B_i^{-1}$ on multiple vectors, this can be done in a single forward elimination and backward substitution. More details on the solution of linear systems with multiple right-hand sides are given in Section \ref{sec:numericalresults}.
The preconditioner ${M}^{-1}_{\text{ORAS}}$ \eqref{eq:ORAS} is naturally parallel since its assembly requires the concurrent factorization of each $\{B_i\}_{1 \leqslant i \leqslant N_S}$, which are typically stored locally on different processes in a distributed computing context. Likewise, applying \eqref{eq:ORAS} to a distributed vector only requires peer-to-peer communications between neighboring subdomains, and a local forward elimination and backward substitution. See chapter~8 of~\cite{Dolean:2015:IDD} for a more detailed analysis.

\subsection{Software stack}
%but our framework interacts with other DSLs such as Feel++ \cite{prud:2012:feel++}. For the computations it is used together with the HPDDM library.
All operators related to the domain decomposition method can be generated using finite element Domain-Specific Languages (DSL). Here we use FreeFem++ \cite{Hecht:2012:NDF} (\url{http://www.freefem.org/ff++/}) since it has already been proven that it can enable large-scale simulations using overlapping Schwarz methods \cite{jolivet:2012:high} when used in combination with the library HPDDM \cite{JolHecNat:2013:hpddm} (High-Performance unified framework for Domain Decomposition Methods, \url{https://github.com/hpddm/hpddm}). 
%\textcolor{corrections}{We use FreeFem++ in combination with the library HPDDM \cite{JolHecNat:2013:hpddm} (High-Performance unified framework for Domain Decomposition Methods, \url{https://github.com/hpddm/hpddm}).} 
HPDDM implements several domain decomposition methods such as RAS, ORAS, FETI, and BNN. It uses multiple levels of parallelism: communication between subdomains is based on the Message Passing Interface (MPI), and computations in the subdomains can be executed on several threads by calling optimized BLAS libraries (such as Intel MKL), or shared-memory direct solvers like PARDISO. Domain decomposition methods naturally offer good parallel properties on distributed architectures. The computational domain is decomposed into subdomains in which concurrent computations are performed. The coupling between subdomains requires communications between computing nodes via messages.
The strong scalability of the ORAS preconditioner as implemented in HPDDM for the direct problem presented in Section~\ref{sec:direct} will be assessed in Section~\ref{sec:numericalresults}.

\subsection{Partition of unity}

\textcolor{corrections}{Here we describe the construction of the partition of unity~\eqref{eq:algebPartUnity} in more details, as its construction in the context of N\'ed\'elec edge elements is non-trivial.}

%\textcolor{corrections}{All the operators related to the domain decomposition method can be generated using finite element Domain-Specific Languages (DSL): here we use FreeFem++ (\url{http://www.freefem.org/ff++/}) \cite{Hecht:2012:NDF}.   
%In particular, the construction of the partition of unity is intricate for N\'ed\'elec edge finite elements and to this purpose new tools had to be added to FreeFem++.}

%\textcolor{corrections}{The starting point is the construction of partition of unity functions $\{\chi_i\}_{1\leqslant i \leqslant N_S}$ for the classical P1 linear nodal finite element such that $\sum_{i = 1}^{N_S} \chi_i = 1$. Another requirement here is that not only the function $\chi_i$ but also its derivative are equal to zero on the border of $\mathcal{T}_i^{\delta}$, for each $i=1,\dots,N_S$. } 

The starting point is the construction of partition of unity functions $\{\chi_i\}_{1\leqslant i \leqslant N_S}$ for the classical P1 linear nodal finite element, whose degrees of freedom are the values at the nodes of the mesh.
First of all, we define for $i=1,\dots,N_S$ the function $\widetilde{\chi}_i$ as the continuous piecewise linear function on $\mathcal{T}$, with support contained in $\mathcal{T}_i^{\delta}$, such that 
\begin{equation*}
\widetilde{\chi}_i = 
\begin{cases}
1 & \text{at all nodes of } \mathcal{T}_i^0, \\ 
0 &\text {at all nodes of } \mathcal{T}_i^{\delta} \setminus  \mathcal{T}_i^{0}.
\end{cases}
\end{equation*}
%Let us denote by $\varphi[]$ the vector of degrees of freedom of a function $\varphi$ belonging to a given finite element space.
The function $\chi_i$ can then be defined as the continuous piecewise linear function on $\mathcal{T}$, with support contained in $\mathcal{T}_i^{\delta}$, such that its (discrete) value for each degree of freedom is evaluated by:
\begin{equation}
\label{eq:chi}
\chi_i = \dfrac{\widetilde{\chi}_i}{\displaystyle\sum_{j=1}^{N_S} {\widetilde{\chi}_j} }. %\big|_{ \mathcal{T}^\delta_i \cap  \mathcal{T}^\delta_j}} \,,
\end{equation}
Thus, we have $\sum_{i = 1}^{N_S} \chi_i = \mathbf{1}$ both at the discrete and continuous level.
Remark that if $\delta >1$, not only the function $\chi_i$ but also its derivative is equal to zero on the border of $\mathcal{T}_i^{\delta}$. 
This is essential for a good convergence if \textcolor{corrections}{Robin-type boundary conditions (such as impedance boundary conditions)}  are chosen as transmission conditions at the interfaces between subdomains. Indeed, if this property is satisfied, the continuous version of the ORAS algorithm is equivalent to P.~L.~Lions' algorithm (see~\cite{StCyr:2007:OMA} and \cite{Dolean:2015:IDD}~\S2.3.2). 
Note that in the practical implementation, the functions $\widetilde{\chi}_i$ and $\chi_i$ are constructed locally on $\mathcal{T}_i^{\delta}$, the relevant contribution of the $\widetilde{\chi}_j$ in~\eqref{eq:chi} being on $\mathcal{T}^\delta_j \cap  \mathcal{T}^\delta_i$. This removes all dependency on the global mesh $\mathcal{T}$, which could be otherwise problematic at large scales.

Now, the degrees of freedom of N\'ed\'elec finite elements are associated with the edges of the mesh. For these finite elements, we can build a geometric partition of unity based on the support of the degrees of freedom (the edges of the mesh): the entries of the diagonal matrices $D_i$, $i=1,\dots,N_S$ are obtained for each degree of freedom by interpolating the piecewise linear function $\chi_i$ at the midpoint of the corresponding edge. The partition of unity property~\eqref{eq:algebPartUnity} is then satisfied since $\sum_{i = 1}^{N_S} \chi_i = {1}$.

\textcolor{corrections}{This interpolation is obtained thanks to a new FreeFem++ {scalar} finite element space, which has only the interpolation operator (defined by one quadrature point on each edge) and no basis functions. 
This auxiliary finite element space is available by loading the plugin \texttt{Element\_Mixte3d} and is called \texttt{Edge03ds0}, when using the lowest order edge finite elements (\texttt{Edge03d}) to discretize the problem. 
The example script \texttt{maxwell-3d.edp}, available in the directory \texttt{examples++-hpddm} of the FreeFem++ distribution, shows how to use these new tools for domain decomposition with edge finite elements.}

\section{Computing the scattering parameters}
\label{sec:scattering}

In order to compute the numerical counterparts of the \emph{reflection and transmission coefficients} obtained by the measurement apparatus of the imaging chamber shown in Figure \ref{fig:chamber}, we use the following formula, which is appropriate in the case of open-ended waveguides:
\begin{equation}
\label{sij}
S_{ij} = \frac{\displaystyle\int_{\Gamma_i} \overline{\vE_j} \cdot \vE_i^0}{\displaystyle\int_{\Gamma_i} |\vE_i^0|^2 },  \quad i, j = 1,\dots,N,
\end{equation}
where $\vE_i^0$ is the TE$_{10}$ fundamental mode of the $i$-th receiving waveguide and $\mathbf{E}_j$ is the solution of the problem where the $j$-th waveguide transmits the signal ($\overline{\mathbf{E}_j}$ denotes the complex conjugate of $\mathbf{E}_j$).
The $S_{ij}$ with $i \neq j$ are the transmission coefficients, and the $S_{\!jj}$ are the reflection coefficients.
They are gathered in the scattering matrix, also called \emph{S-matrix}. 

Here we compare the coefficients computed from the simulation with a set of measurements obtained by EMTensor. For this test case, the imaging chamber was filled with a homogeneous matching solution. The electric permittivity $\varepsilon$ of the matching solution is chosen by EMTensor in order to minimize contrasts with the ceramic-loaded waveguides and with the different brain tissues. The choice of the conductivity $\sigma$ of the matching solution is a compromise between the minimization of reflection artifacts from metallic boundaries and the desire to have best possible signal-to-noise ratio. Here the relative complex permittivity of the matching solution at frequency $f = $ \SI{1}{\giga\hertz} is $\varepsilon_r^{\text{gel}} = 44 - 20\mathtt{i}$. The relative complex permittivity inside the ceramic-loaded waveguides is $\varepsilon_r^{\text{cer}} = 59$.

\begin{figure}[h!]
\centering 
\includegraphics[width=0.5\textwidth]{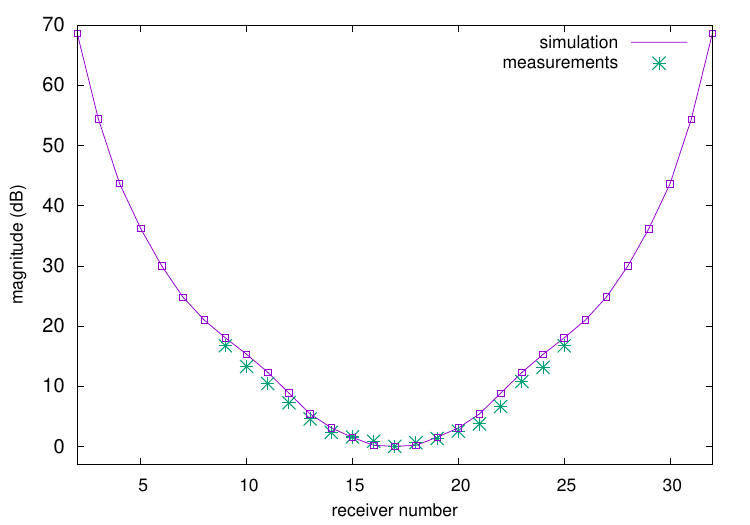} \\
\hspace{-0.35cm}
\includegraphics[width=0.525\textwidth]{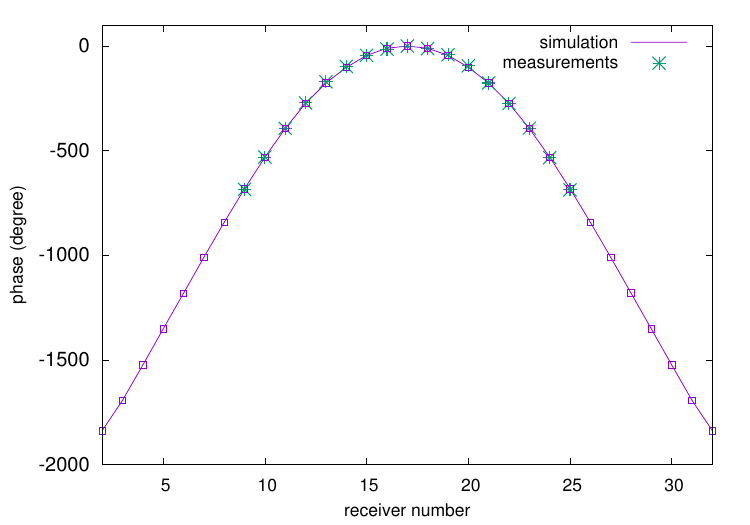} 
\caption{The normalized magnitude (top) and phase (bottom) of the transmission coefficients computed with the simulation and measured experimentally.} 
\label{fig:cmp}
\end{figure}
The set of experimental data at hand given by EMTensor consists in transmission coefficients for transmitting antennas in the second ring from the top.
Figure~\ref{fig:cmp} shows the normalized magnitude (dB) and phase (degree) of the complex coefficients $S_{ij}$ corresponding to a transmitting antenna in the second ring from the top and to the $31$ receiving antennas in the middle ring (note that measured coefficients are available only for $17$ receiving antennas). The magnitude in dB is calculated as $20 \log_{10} (|S_{ij} |)$. The normalization is done by dividing every transmission coefficient by the transmission coefficient corresponding to the receiving antenna directly opposite to the transmitting antenna, which is thus set to $1$. Since we normalize with respect to the coefficient having the lowest expected magnitude, the magnitude of the transmission coefficients displayed in Figure~\ref{fig:cmp} is larger than $0$ dB.
We can see that the transmission coefficients computed from the simulation are in very good agreement with the measurements.

\section{The inverse problem}
\label{sec:inverseproblem}

The inverse problem that we consider consists in finding the unknown dielectric permittivity $\eps(\bf x)$ and conductivity $\sigma(\bf x)$ in $\Omega$, such that the solutions $\vE_j, j = 1,\dots,N$ of problem (\ref{pbdirect}) lead to corresponding scattering parameters $S_{ij}$~(\ref{sij}) that coincide with the measured scattering parameters $S^{\text{mes}}_{ij}$, for $i,j = 1,\dots,N$. In the following, we present the inverse problem in the continuous setting for clarity.

Let $\kappa = \mu_0 (\omega^2 \eps - \text{i} \omega \sigma)$ be the unknown complex parameter of our inverse problem, and let us denote by $\vE_j(\kappa)$ the solution of the direct problem (\ref{pbdirect}) with dielectric permittivity $\eps$ and conductivity $\sigma$. The corresponding scattering parameters will be denoted by $S_{ij}(\kappa)$ for $i,j = 1,\dots,N$:
\begin{equation*}
% \label{sijkappa}
S_{ij}(\kappa) = \frac{\displaystyle\int_{\Gamma_i} \overline{\vE_j(\kappa)} \cdot \vE_i^0 }{\displaystyle\int_{\Gamma_i} |\vE_i^0|^2 },  \quad i, j = 1,\dots,N.
\end{equation*}
The misfit of the parameter $\kappa$ to the data can be defined through the following functional:
\begin{equation}
\label{defJ}
J(\kappa) = \frac{1}{2} \sum_{j=1}^{N}\sum_{i=1}^{N}  \left| S_{ij}(\kappa)  - S_{ij}^{\text{mes}}\right|^2 = \frac{1}{2} \sum_{j=1}^{N}\sum_{i=1}^{N}  \left| \frac{\displaystyle\int_{\Gamma_i} \overline{\vE_j(\kappa)} \cdot \vE_i^0 }{\displaystyle\int_{\Gamma_i} |\vE_i^0|^2 }  - S_{ij}^{\text{mes}}\right|^2.
\end{equation}
In a classical way, solving the inverse problem then consists in minimizing the functional $J$ with respect to the parameter $\kappa$. Computing the differential of $J$ in a given arbitrary direction $\delta\kappa$ yields
\begin{equation*}
DJ (\kappa, \delta \kappa) = \sum_{j=1}^{N}\sum_{i=1}^{N} \Re \left[\overline{\left(S_{ij}(\kappa) - S^{\text{mes}}_{ij}\right)}  \frac{\displaystyle \int_{\Gamma_i} \overline{\delta\vE_j(\kappa)} \cdot \vE_i^0 }{\displaystyle \int_{\Gamma_i} |\vE_i^0|^2 } \right], \quad \delta \kappa \in \mathbb{C},
\end{equation*}
where $\delta\vE_j(\kappa)$ is the solution of the following linearized problem: 
\begin{equation}
\quad \left\{
\label{pblin}
\begin{aligned}
\nabla \times (\nabla \times \delta\vE_j) -  \kappa \delta\vE_j &= \delta\kappa \vE_j &&\mbox{ in } \Omega, \\
\delta\vE_j \times \vn&= {0}  &&\mbox{ on } {\color{black}\Gamma_\text{m}}, \\
(\nabla \times \delta\vE_j) \times \vn +\text{i} \beta  \vn \times (\delta\vE_j \times \vn) &= {0} &&\mbox{ on } {\color{black}\Gamma_i} \mbox{ , } i = 1,\dots,N.
\end{aligned}
\right.
\end{equation}

We now use the adjoint approach in order to simplify the expression of $DJ$. This will allow us to compute the gradient efficiently after discretization, with a number of computations independent of the size of the parameter space. Considering the variational formulation of problem (\ref{pblin}) with a test function $\vF$ and integrating by parts, we get
\begin{equation*}
 \begin{aligned}
\int_\Omega \delta \kappa \vE_j \cdot \vF &=\int_\Omega \left(\nabla \times ( \nabla \times \delta \vE_j) - \kappa \delta \vE_j \right) \cdot \vF  \\
&= \int_\Omega \left(\nabla \times ( \nabla \times \vF) - \kappa \vF \right) \cdot \delta \vE_j  - \int_{\partial \Omega} ((\nabla \times \delta \vE_j) \times \vn)\cdot \vF\ 
+ \int_{\partial \Omega} ((\nabla \times \vF)\times \vn) \cdot \delta \vE_j \\
&=  \int_\Omega \left(\nabla \times ( \nabla \times \vF) - \kappa \vF \right) \cdot \delta \vE_j + \sum_{i=1}^{N} \int_{\Gamma_i} \text{i} \beta (\vn \times (\vF \times \vn) ) \cdot \delta \vE_j  \\
&+ \int_{\Gamma_\text{m}}  (\nabla \times \delta \vE_j) \cdot (\vF \times \vn) + \sum_{i=1}^{N} \int_{\Gamma_i} ((\nabla \times \vF) \times \vn)\cdot  \delta \vE_j .
\end{aligned}
\end{equation*}
Introducing the solution $\vF_j(\kappa)$ of the following adjoint problem
\begin{equation}
\quad \left\{
\label{pbadj}
\begin{aligned}
\nabla \times (\nabla \times \vF_j) -  \kappa \vF_j &= {0}   &&\mbox{ in } \Omega, \\
\vF_j \times \vn&= {0}  &&\mbox{ on } {\color{black}\Gamma_\text{m}}, \\
(\nabla \times \vF_j) \times \vn +\text{i} \beta \vn \times (\vF_j \times \vn) &= \frac{(S_{ij}(\kappa)  - S_{ij}^{\text{mes}})}{\displaystyle\int_{\Gamma_i} | \vE_i^0 |^2 } \overline{\vE_i^0} &&\mbox{ on } {\color{black}\Gamma_i} \mbox{ , } i = 1,\dots,N,
\end{aligned}
\right.
\end{equation}
we get
\begin{equation*}
 \int_\Omega \delta \kappa \vE_j \cdot \vF_j = \sum_{i=1}^{N} (S_{ij}(\kappa)  - S_{ij}^{\text{mes}}) \frac{\displaystyle\int_{\Gamma_i} \overline{\vE_i^0} \cdot \delta \vE_j}{\displaystyle\int_{\Gamma_i} | \vE_i^0 |^2 } .
\end{equation*}

Finally, the differential of $J$ can be computed as
\begin{equation*}
% \label{gradientadjoint}
DJ (\kappa, \delta \kappa) =  \Re \left[\int_\Omega {\overline{\delta \kappa}\, \sum_{j=1}^{N} \overline{\vE_j \cdot \vF_j}} \right] .
\end{equation*}

%The numerical results presented in Section \ref{sec:numericalresults} are obtained using a nonlinear conjugate gradient algorithm with cubic line search based on the Armijo--Goldstein conditions.
We can then compute the gradient to use in a gradient-based local optimization algorithm. The numerical results presented in Section \ref{sec:numericalresults} are obtained using a limited-memory Broyden-Fletcher-Goldfarb-Shanno (L-BFGS) algorithm. Note that every evaluation of $J$ requires the solution of the state problem~\eqref{pbdirect} while the computation of the gradient requires the solution of~\eqref{pbdirect} as well as the solution of the adjoint problem~\eqref{pbadj}. Moreover, the state and adjoint problems use the same operator. Therefore, the computation of the gradient only needs the assembly of one matrix and its associated domain decomposition preconditioner.

Numerical results for the reconstruction of a hemorrhagic stroke from synthetic data are presented in the next section. The functional $J$ considered in the numerical results is slightly different from~\eqref{defJ}, as we add a normalization term for each pair $(i,j)$ as well as a Tikhonov regularizing term:
\begin{equation}
\label{defJ2}
J(\kappa) = \frac{1}{2} \sum_{j=1}^{N}\sum_{i=1}^{N} \frac{\bigl \lvert S_{ij}(\kappa)  - S_{ij}^{\text{mes}}\bigr \rvert^2}{\bigl \lvert S_{ij}^{\text{empty}}\bigr \rvert^2} + \frac{\alpha}{2} \int_\Omega |\nabla \kappa|^2,
\end{equation}
where $S_{ij}^{\text{empty}}$ refers to the coefficients computed from the simulation with the empty chamber, that is the chamber filled only with the homogeneous matching solution as described in the previous section, with no object inside. In this way, the contribution of each pair $(i,j)$ in the misfit functional is normalized and does not depend on the amplitude of the coefficient, which can vary greatly between pairs $(i,j)$ as displayed in Figure~\ref{fig:cmp}. The Tikhonov regularizing term aims at reducing the effects of noise in the data. For now, the regularization parameter $\alpha$ is chosen empirically so as to obtain a visually good compromise between reducing the effects of noise and keeping the reconstructed image pertinent. All calculations carried out in this section can be accommodated in a straightforward manner to definition~\eqref{defJ2} of the functional.

\section{Numerical results}
\label{sec:numericalresults}

Results in this paper were obtained on Curie, a system composed of 5,040 nodes made of two eight-core Intel Sandy Bridge processors clocked at \SI{2.7}{\giga\hertz}. The interconnect is an InfiniBand QDR full fat tree and the MPI implementation used was BullxMPI version 1.2.8.4. Intel compilers and Math Kernel Library in their version 16.0.2.181 were used for all binaries and shared libraries, and as the linear algebra backend for dense computations. One-level preconditioners such as \eqref{eq:ORAS}\textcolor{corrections}{, whose action on a vector is implemented by} 
%assembled by 
HPDDM, require the use of a sparse direct solver. In the following experiments, we have been using either PARDISO \cite{schenk2004solving} from Intel MKL or MUMPS \cite{amestoy2001fully}.
All linear systems resulting from the edge finite elements discretization are solved by GMRES right-preconditioned with ORAS~\eqref{eq:ORAS} as implemented in HPDDM. The GMRES algorithm is stopped once the unpreconditioned relative residual is lower than~$10^{-8}$.
First, we perform a strong scaling analysis in order to assess the efficiency of our preconditioner. Then, we assess the feasibility of the microwave imaging technique presented in this paper for stroke detection and monitoring through a numerical example in a realistic configuration. We use synthetic data corresponding to a numerical model of a virtual human head with a simulated hemorrhagic stroke as input for the inverse problem.

\subsection{Scaling analysis}
\label{sec:scaling}

\begin{figure}%[h!]
\centering
\includegraphics{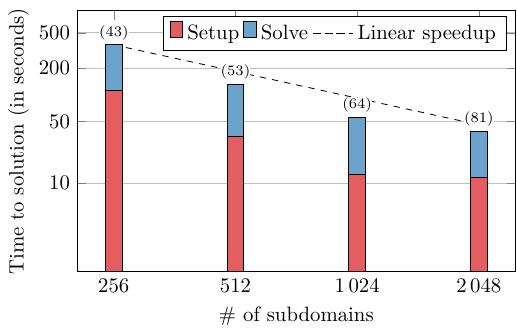} 
\caption{Strong scaling experiment. Colors indicate the fraction of the total time spent in the setup and solution phases. The number of GMRES iterations is reported in parentheses.}
\label{figscaling}
\end{figure}
\begin{table}%[h!]
\centering
\pgfplotstableread{
X       Zn            Pb               It
256     2.9336e+02    7.3061e+01       43
512     9.5107e+01    3.6922e+01       53
1024    3.5130e+01    2.0549e+01       64
2048    2.5885e+01    1.2772e+01       81
}\datatable
\pgfplotstablegetelem{0}{X}\of{\datatable}
\let\relativeTo\pgfplotsretval
\pgfplotstablecreatecol[create col/expr={\thisrow{Zn} +\thisrow{Pb}}]{sum}\datatable
\pgfplotstablegetelem{0}{sum}\of{\datatable}
\let\sum\pgfplotsretval
\pgfplotstablecreatecol[create col/expr={\sum/\thisrow{sum}}]{speedup}\datatable
\pgfplotstabletypeset[columns={X,Zn,Pb,It,speedup},columns/X/.style = {column name =$N_S$},columns/Zn/.style = {column name =Setup,dec sep align},columns/Pb/.style = {column name =Solve},columns/It/.style = {column name =\# of iterations},columns/speedup/.style = {column name =Speedup,precision =1},
every head row/.style={
    % as in the previous example, this patches the first row:
    before row={\hline
    },
    after row=\hline
},every last row/.style={after row=\hline},
every first column/.style={
    column type/.add={|}{},
},
every last column/.style={
    column type/.add={}{|},
}
]{\datatable}

\caption{Strong scaling experiment. Timings (in seconds) of the setup and solution phases.}
\label{tabscaling}
\end{table}
Using the domain decomposition preconditioner~\eqref{eq:ORAS}, we solve the direct problem corresponding to the setting of Section~\ref{sec:scattering} where the chamber is filled with a homogeneous matching solution. We consider a right-hand side corresponding to a transmitting antenna in the second ring from the top. Given a fine mesh of the domain composed of $82$ million tetrahedra, we increase the number of MPI processes to solve the linear system of $96$ million double-precision complex unknowns yielded by the discretization of Maxwell's equation using edge elements. The global unstructured mesh is partitioned using SCOTCH \cite{pellegrini1996scotch} and the local solver is PARDISO from Intel MKL. We use one subdomain and two OpenMP threads per MPI process. Results are reported in Table~\ref{tabscaling} and illustrated in Figure~\ref{figscaling} with a plot of the time to solution including both the setup and solution phases on $256$ up to $2048$ subdomains. 
%The setup phase time is the maximum over all the subdomains, $i=1,\dots,N_S$, of the time for the factorization of the local subproblem matrices $B_i$ in~\eqref{eq:ORAS}, while the solution phase time is the time to solve the linear system with GMRES. 
The setup time corresponds to the maximum time spent for the factorization of the local subproblem matrix $B_i$ in~\eqref{eq:ORAS} over all subdomains, while the solution time corresponds to the time needed to solve the linear system with GMRES.
We are able to obtain very good speedups up to $4096$ cores ($2048$ subdomains) on Curie, with a superlinear speedup of $9.5$ between $256$ and $2048$ subdomains. \textcolor{corrections}{Indeed, for the range of process counts we are considering here, the cost of the setup (performing exact $LDL^H$ factorizations in subdomains) is greater than the one of the solution phase. Moreover, since the cost of computing such factorizations decreases quadratically with respect to the size of the local problems, it is possible to achieve superlinear speedups in the strong-scaling regime.}

\color{corrections}

\subsection{Direct simulation of a hemorrhagic stroke using a virtual head model}

\begin{figure}[h!]
\centering 
\includegraphics[height = .3\textwidth]{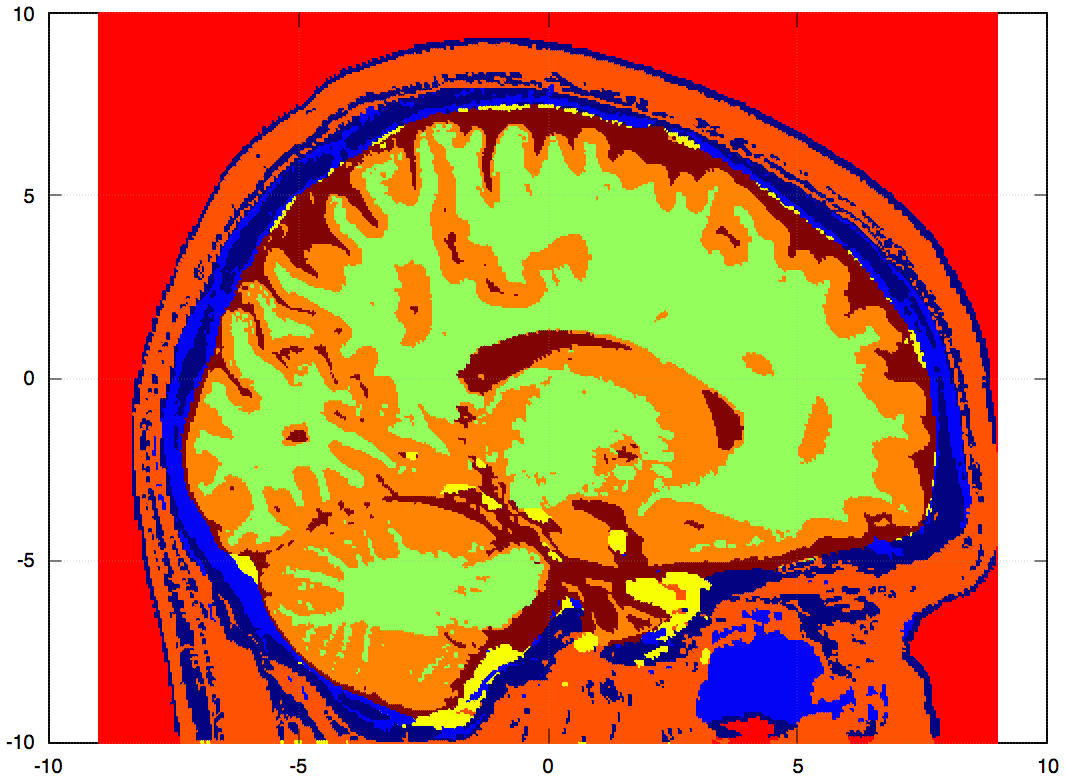} \quad
\includegraphics[height = .28\textwidth,trim = 0cm -1cm 0cm 3cm]{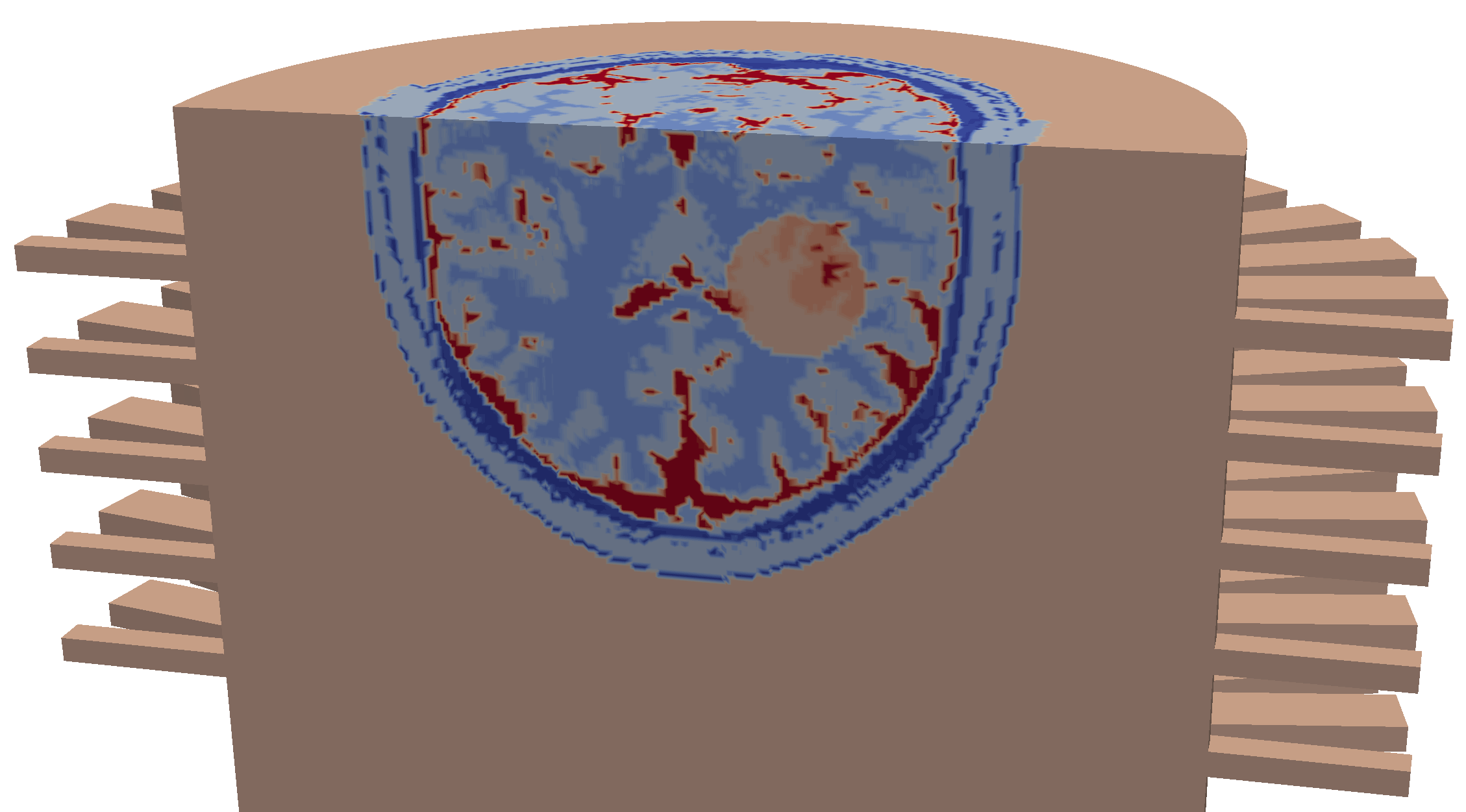}
\caption{Left: sagittal section of the brain. Right: numerical head immersed in the imaging chamber, with a simulated ellipsoid-shaped hemorrhagic stroke.}
\label{figbrain}
\end{figure}

The numerical model of the virtual head comes from CT and MRI tomographic images and consists of a complex permittivity map of $362\times 434\times 362$ data points. Figure \ref{figbrain} (left) shows a sagittal section of the head. In the simulation, the head is immersed in the imaging chamber as shown in Figure \ref{figbrain} (right). In order to simulate a hemorrhagic stroke, a synthetic stroke is added in the form of an ellipsoid in which the value of the complex permittivity $\varepsilon_r$ has been increased. For this test case, the value of the permittivity in the ellipsoid is taken as the mean value between the relative permittivity of the original healthy brain and the relative permittivity of blood at frequency $f = $ \SI{1}{\giga\hertz}, $\varepsilon_r^{\text{blood}} = 68 - 44 \text{i}$. The imaging chamber is filled with a matching solution. The relative permittivity of the matching solution is chosen by EMTensor as explained in Section \ref{sec:scattering} and is equal to $\varepsilon_r^{\text{gel}} = 44 - 20 \text{i}$ at frequency $f = $ \SI{1}{\giga\hertz}. In the real setting, a special membrane fitting the shape of the head is used in order to isolate the head from the matching medium. We do not take this membrane into account in this synthetic test case.
The synthetic data are obtained by solving the direct problem on a mesh composed of $17.6$ million tetrahedra (corresponding to approximately $20$ points per wavelength) and consist in the transmission and reflection coefficients $S_{ij}$ calculated from the simulated electric field as in~\eqref{sij}.

\begin{figure}[h!]
\centering 
\includegraphics[width = .3\textwidth]{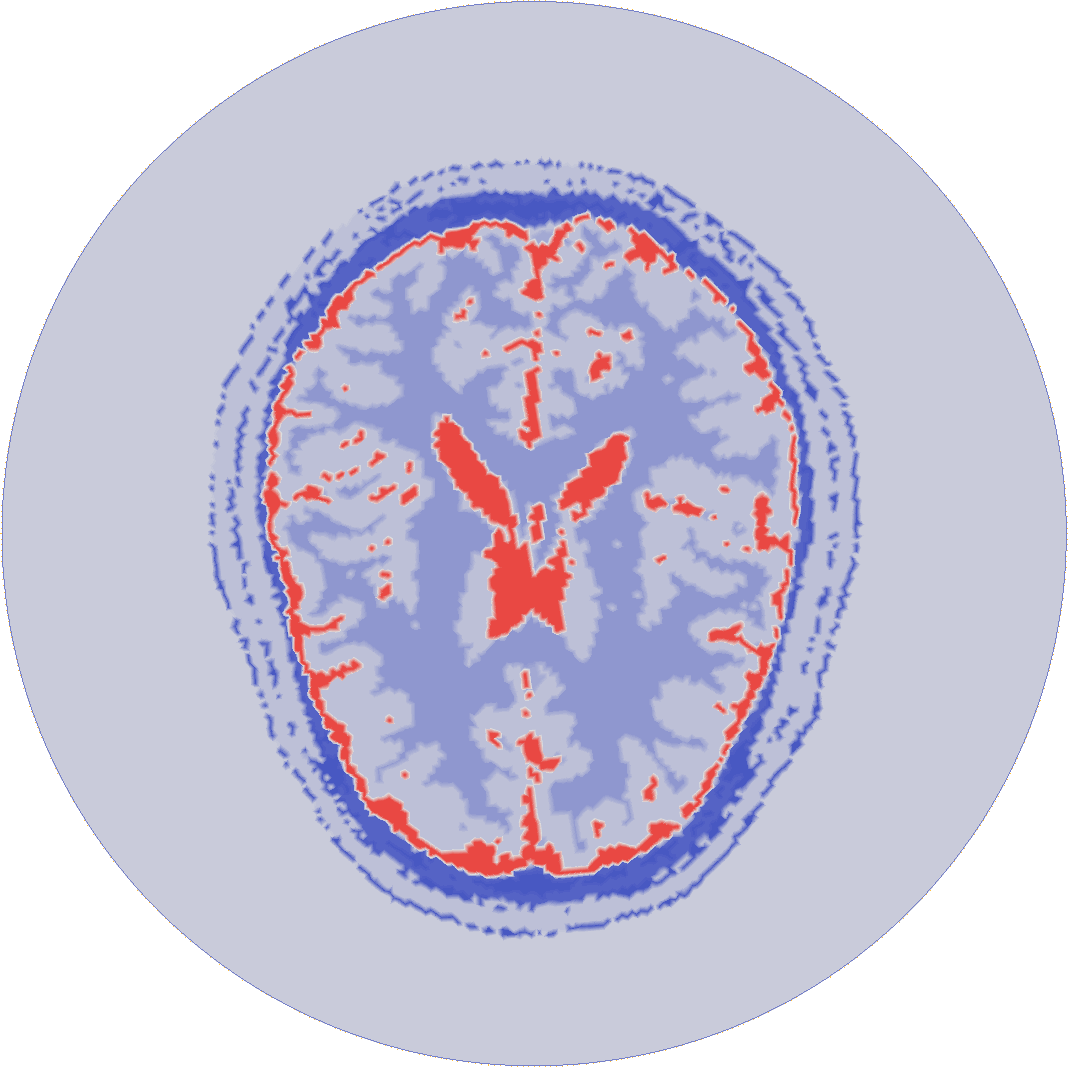}
\includegraphics[width = .3\textwidth]{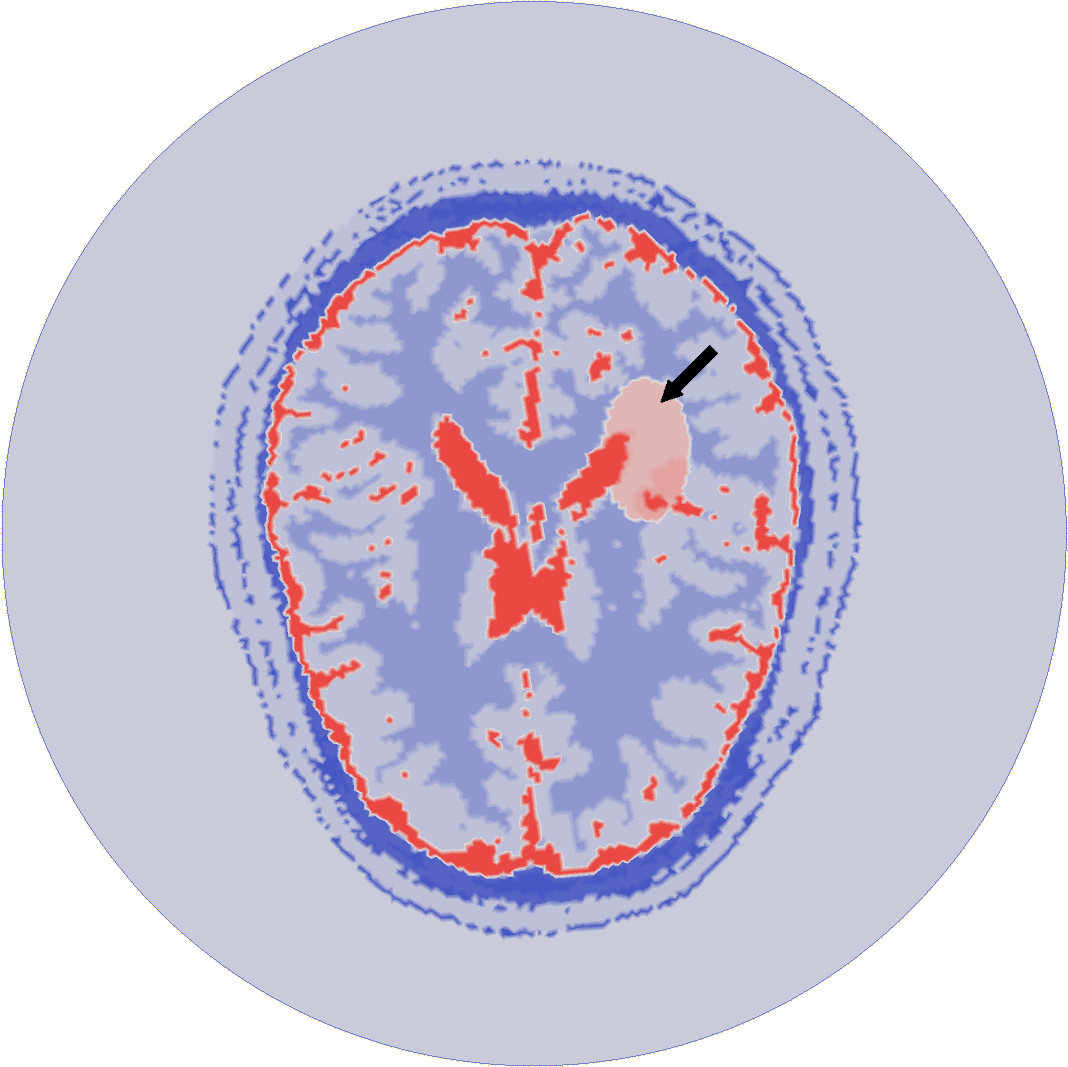}
\includegraphics[width = .3\textwidth]{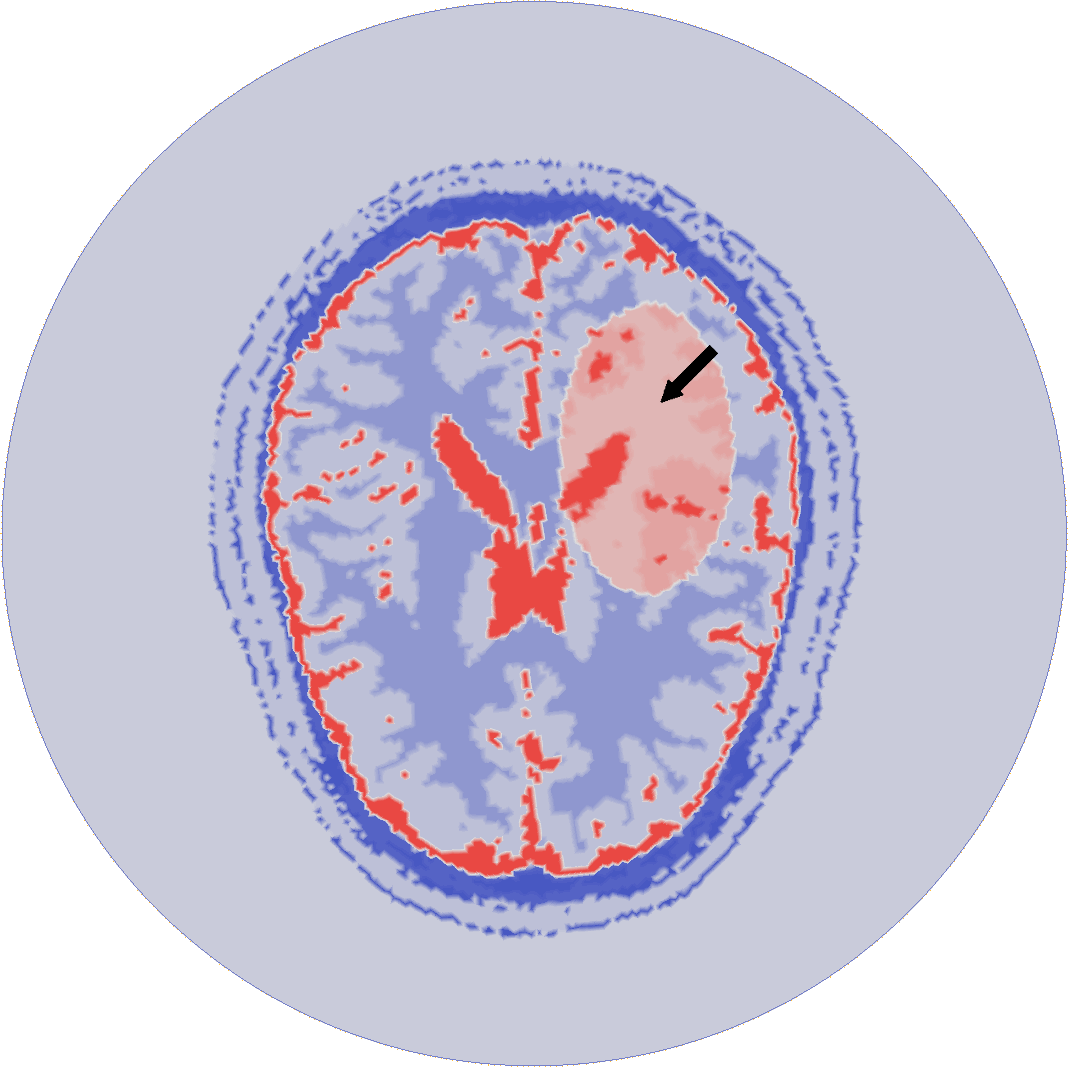}
\caption{Imaginary part of the exact permittivity for a time evolution of a synthetic hemorrhagic stroke. From left to right: healthy brain, small stroke, large stroke. The size of the ellipsoid is $\SI{3.9}{\centi\meter} \times \SI{2.3}{\centi\meter} \times \SI{2.3}{\centi\meter}$ and $\SI{7.7}{\centi\meter} \times \SI{4.6}{\centi\meter} \times \SI{4.6}{\centi\meter}$ in the middle and right picture respectively.}
\label{fig:exact}
\end{figure}

We simulate the evolution of the hemorrhagic stroke by increasing the size of the ellipsoid in which the value of the permittivity is raised. Thus, we solve the direct problem for three different complex permittivity maps, shown in Figure~\ref{fig:exact}: healthy brain, small stroke and large stroke.

\begin{figure}[h!]
\centering 
\includegraphics[width = .49\textwidth]{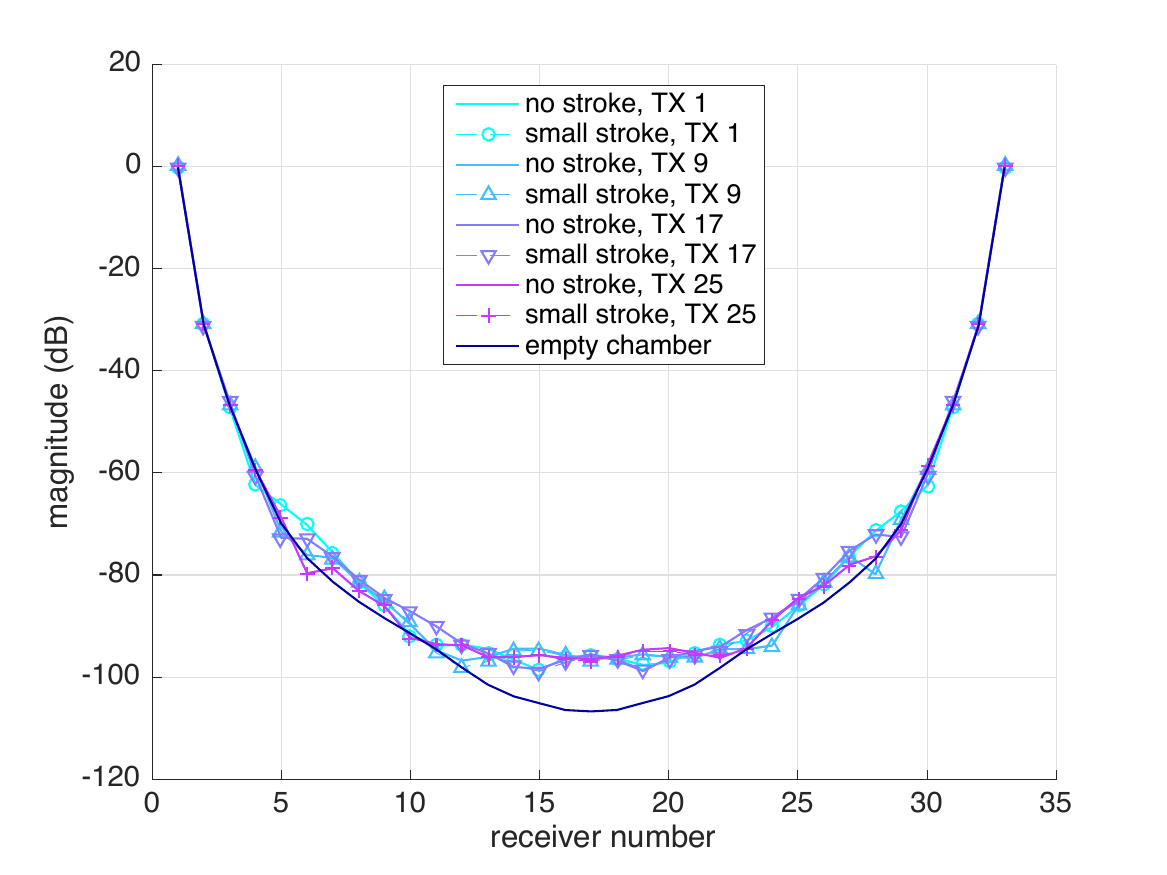}
\includegraphics[width = .49\textwidth]{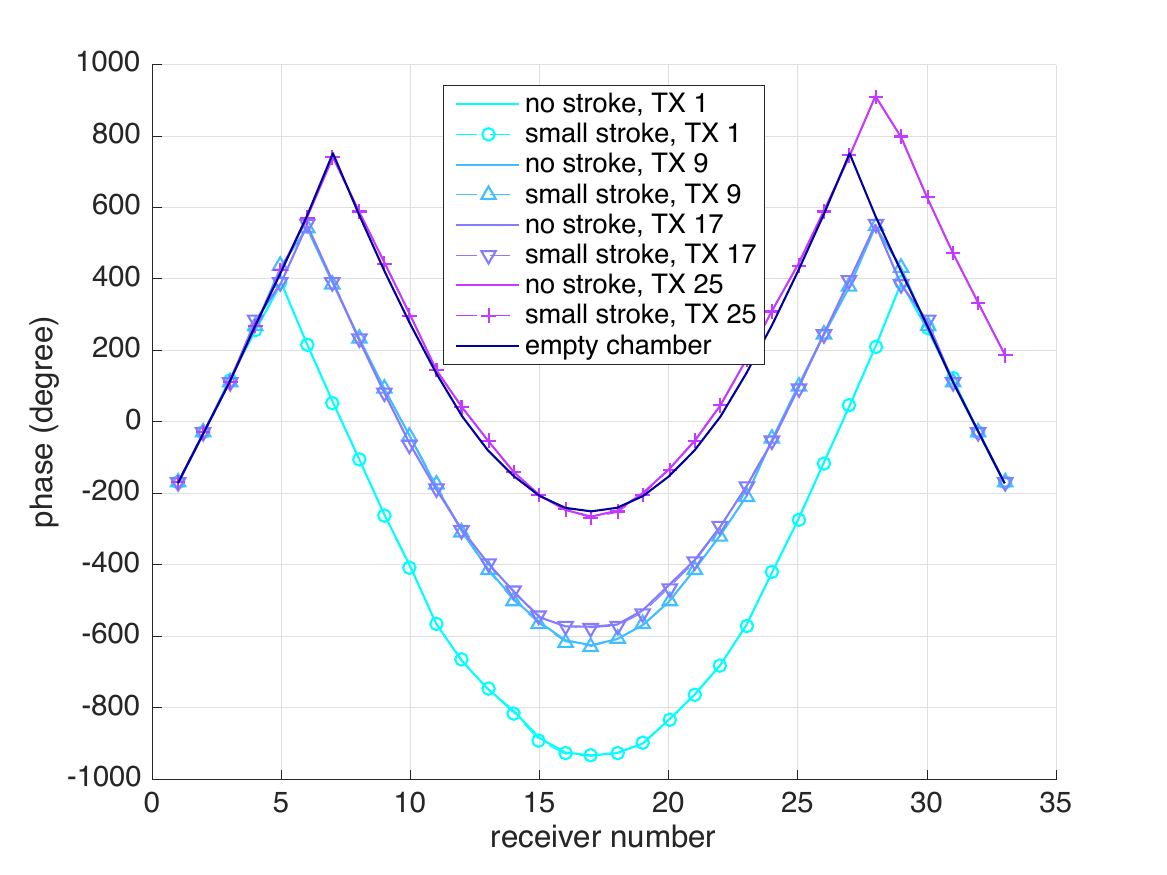}
\includegraphics[width = .49\textwidth]{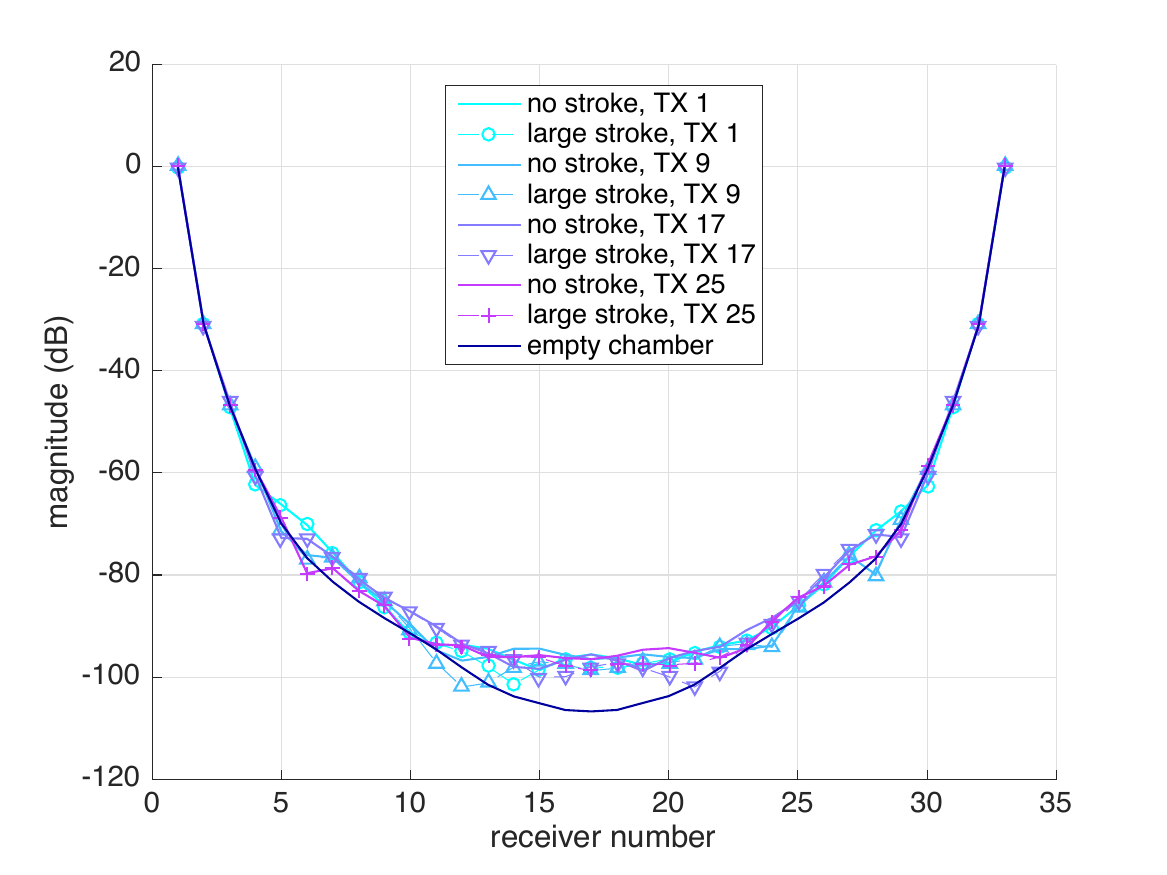}
\includegraphics[width = .49\textwidth]{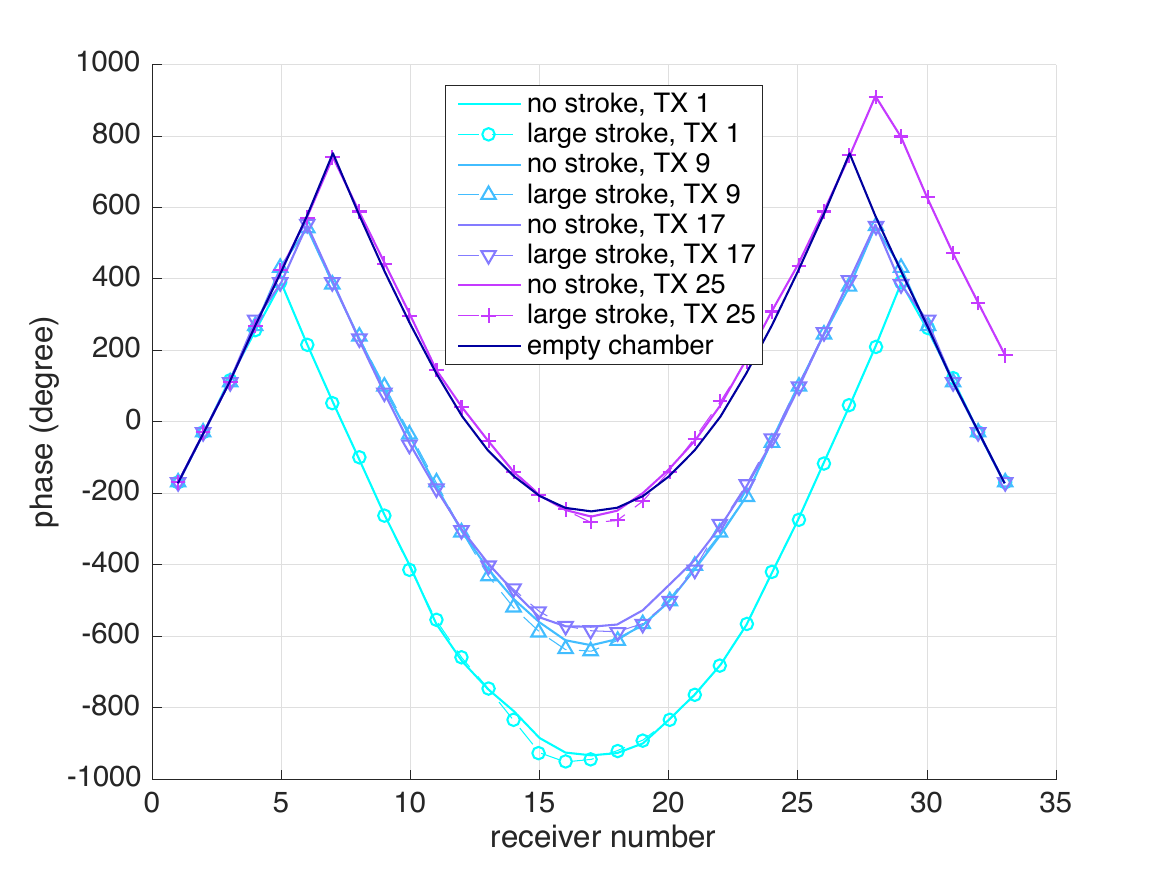}
\caption{Magnitude (left) and phase (right) of the $S_{ij}$ corresponding to transmitting antennas $j=1,9,17$ and $25$ of the top ring, and to all 32 receiving antennas of the top ring $i = 1,\dots,32$. Plain curves represent the scattering parameters for the healthy brain, while dashed curves with symbols correspond to the small (top) and large (bottom) stroke. The scattering parameters for the empty chamber (homogeneous gel) are also represented (dark blue). The receiver numbers have been shifted for each curve so that receiver 17 corresponds to the antenna opposite to the transmitting antenna.}
\label{fig:exactSij}
\end{figure}

Figure~\ref{fig:exactSij} shows the magnitude (left) and phase (right) of the complex coefficients $S_{ij}$ corresponding to transmitting antennas $j=1,9,17$ and $25$ of the top ring, and to all 32 receiving antennas of the top ring $i = 1,\dots,32$. Each transmitting antenna is represented in the curves by a different color. Plain curves represent the scattering parameters for the healthy brain, while dashed curves with symbols correspond to the small (top) and large (bottom) stroke. The scattering parameters for the empty chamber (homogeneous gel) are also represented. The receiver numbers have been shifted for each curve so that receiver 17 corresponds to the antenna opposite to the transmitting antenna.

First, we can see that the dynamic range of the magnitude of the signal (left curves) is very large, from $0$ dB for the reflection coefficients down to $-100$ dB for the transmission coefficients corresponding to the receiver opposite to the transmitting antenna. This highlights the need for a sophisticated electronic measurement device as well as for a high-fidelity simulation tool. We can also see that the farther the receiver (the opposite receiver being the farthest), the more attenuated the signal is, because the wave propagates through more tissues. The scattering parameters corresponding to the farthest receivers thus carry more information about the brain tissues, which is why we see more discrepancies between the healthy brain and the brain with stroke for the opposite receivers. This justifies the normalization by the empty chamber coefficients in the cost functional~\eqref{defJ2} in order to give weight to the information contained in these very attenuated measurements.

\subsection{Reconstruction of a hemorrhagic stroke from synthetic data}

%REF Electromagnetic Tomography for Brain Imaging: initial assessment for stroke detection

The synthetic data obtained by solving the direct problem for the healthy brain, small stroke and large stroke are used as input for the inverse problem. We add noise to the real and imaginary parts of the coefficients $S_{ij}$ (10\% additive Gaussian white noise, with different values for real and imaginary parts). Furthermore, we assume no a priori knowledge on the permittivity inside the chamber, except that we set the initial guess for the inverse problem as the homogeneous matching solution everywhere inside the chamber.

We use a piecewise linear approximation of the unknown parameter $\kappa$, defined on the same mesh used to solve the state and adjoint problems.

%For the purpose of parallel computations, the partitioning introduced by the domain decomposition method is also used to compute and store locally in each subdomain every entity involved in the inverse problem, such as the parameter $\kappa$ and the gradient.

\paragraph{Exposing multiple levels of parallelism} As is usually the case with most medical imaging techniques, the reconstruction is done layer by layer. For the imaging chamber of EMTensor that we study in this paper, one layer corresponds to one of the five rings of $32$ antennas. This allows us to exhibit another level of parallelism, by solving an inverse problem independently for each of the five rings in parallel. More precisely, each of these inverse problems is solved in a domain truncated around the corresponding ring of antennas, containing at most two other rings (one ring above and one ring below). We impose absorbing boundary conditions on the artificial boundaries of the truncated computational domain. For each inverse problem, only the coefficients $S_{ij}$ with transmitting antennas $j$ in the corresponding ring are taken into account: we consider $32$ antennas as transmitters and at most $96$ antennas as receivers.

Moreover, evaluating the functional or its gradient requires the solution of a linear system with $32$ right-hand sides, one right-hand side per transmitter. This introduces a trivial level of parallelism since the solution corresponding to each right-hand side can be computed independently.

We have thus overall three levels of parallelism: independent inverse problems for each layer, domain decomposition and multiple independent right-hand sides.\\

\noindent However when considering a finite number of available processors, there is a tradeoff between the parallelism induced by the multiple right-hand sides and the parallelism induced by the domain decomposition method. Additionally, to give a complete picture of our acceleration techniques, we mention the fact that we solve for multiple right-hand sides simultaneously using a pseudo-block method implemented inside GMRES which consists in fusing the multiple arithmetic operations corresponding to each right-hand side (matrix-vector products, dot products), resulting in higher arithmetic intensity. The scaling behavior of this pseudo-block algorithm with respect to the number of right-hand sides is nonlinear, as is the scaling behavior of the domain decomposition method with respect to the number of subdomains. Thus, for a given number of processors, we find the optimal tradeoff between parallelizing with respect to the number of subdomains or right-hand sides through trial and error. Note that in the real setting envisioned in this work, where parallel computations for fast stroke diagnosis would be offloaded to potential available clusters, finding this optimal tradeoff along with fine-tuning the different run-time parameters, and also optimizing the compilation process, can be done using offline auto-tuning methods. Indeed, only input measurement data would differ between online runs, and this would not affect the run-time behaviour of the algorithm as measurements datasets are comparable. However, we did not consider auto-tuning methods here, as it is beyond the scope of the paper.

%\begin{figure}
%\centering 
%\includegraphics[width = .3\textwidth]{./figures/sainimag_v2.png}
%\includegraphics[width = .3\textwidth]{./figures/avc1imag_v2.png}
%\includegraphics[width = .3\textwidth]{./figures/avc2imag_v2.png}\\
%\includegraphics[width = .3\textwidth]{./figures/numsainimag15PPLO_v2.png}
%\includegraphics[width = .3\textwidth]{./figures/numavc1imag15PPLO_v2.png}
%\includegraphics[width = .3\textwidth]{./figures/numavc2imag15PPLO_v2.png}
%\end{figure}
%\begin{figure}
%\centering 
%\includegraphics[width = .3\textwidth]{./figures/sainimagnew.png}
%\includegraphics[width = .3\textwidth]{./figures/avc1imagnew.png}
%\includegraphics[width = .3\textwidth]{./figures/avc2imagnew.png}\\
%\includegraphics[width = .3\textwidth]{./figures/numsainimagnew.png}
%\includegraphics[width = .3\textwidth]{./figures/numavc1imagnew.png}
%\includegraphics[width = .3\textwidth]{./figures/numavc2imagnew.png}
%\end{figure}
\begin{figure}[h!]
\centering 
\includegraphics[width = .3\textwidth]{./figures/sainimagnew_div.png}
\includegraphics[width = .3\textwidth]{./figures/avc1imagnew_div.png}
\includegraphics[width = .3\textwidth]{./figures/avc2imagnew_div.png}\\
\includegraphics[width = .3\textwidth]{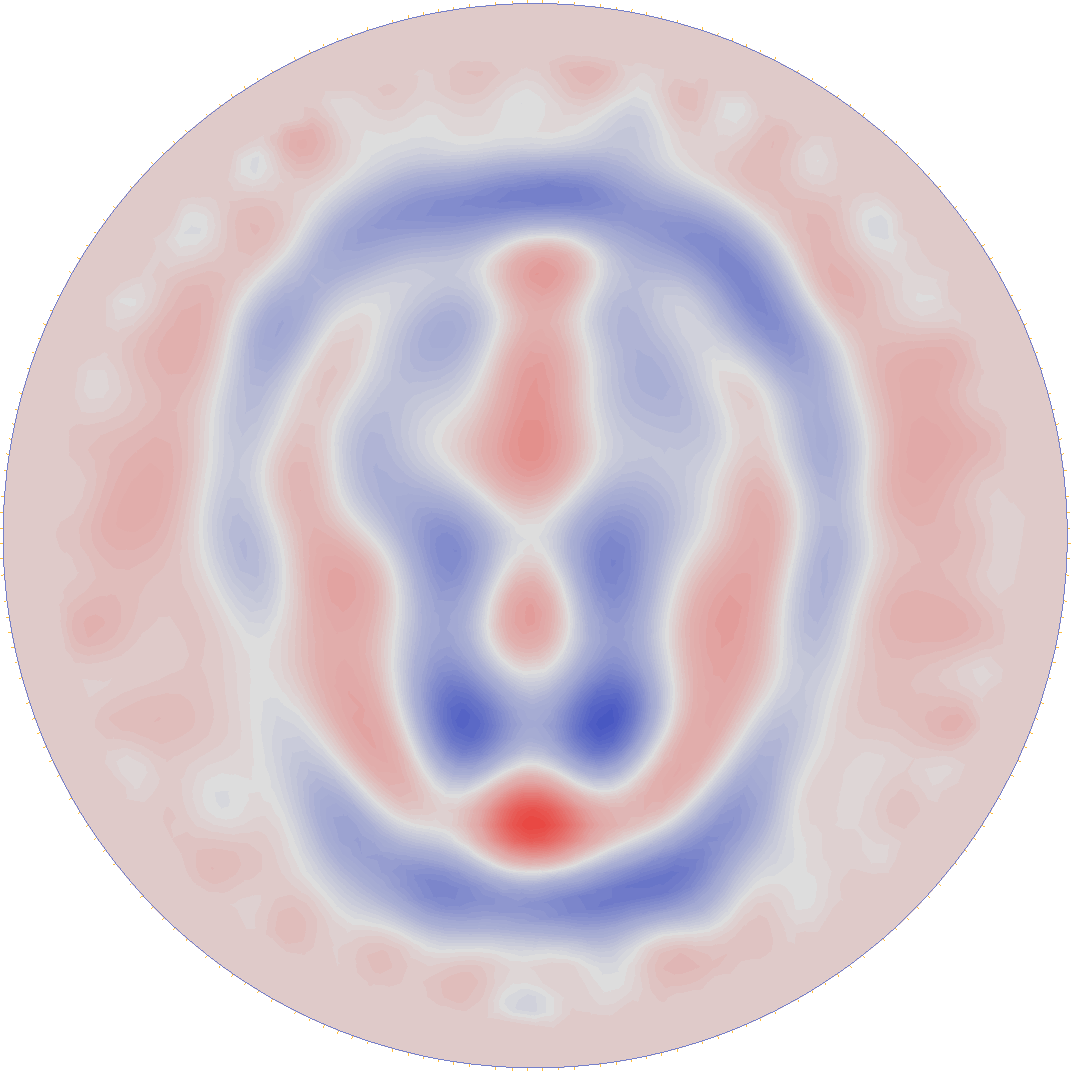}
\includegraphics[width = .3\textwidth]{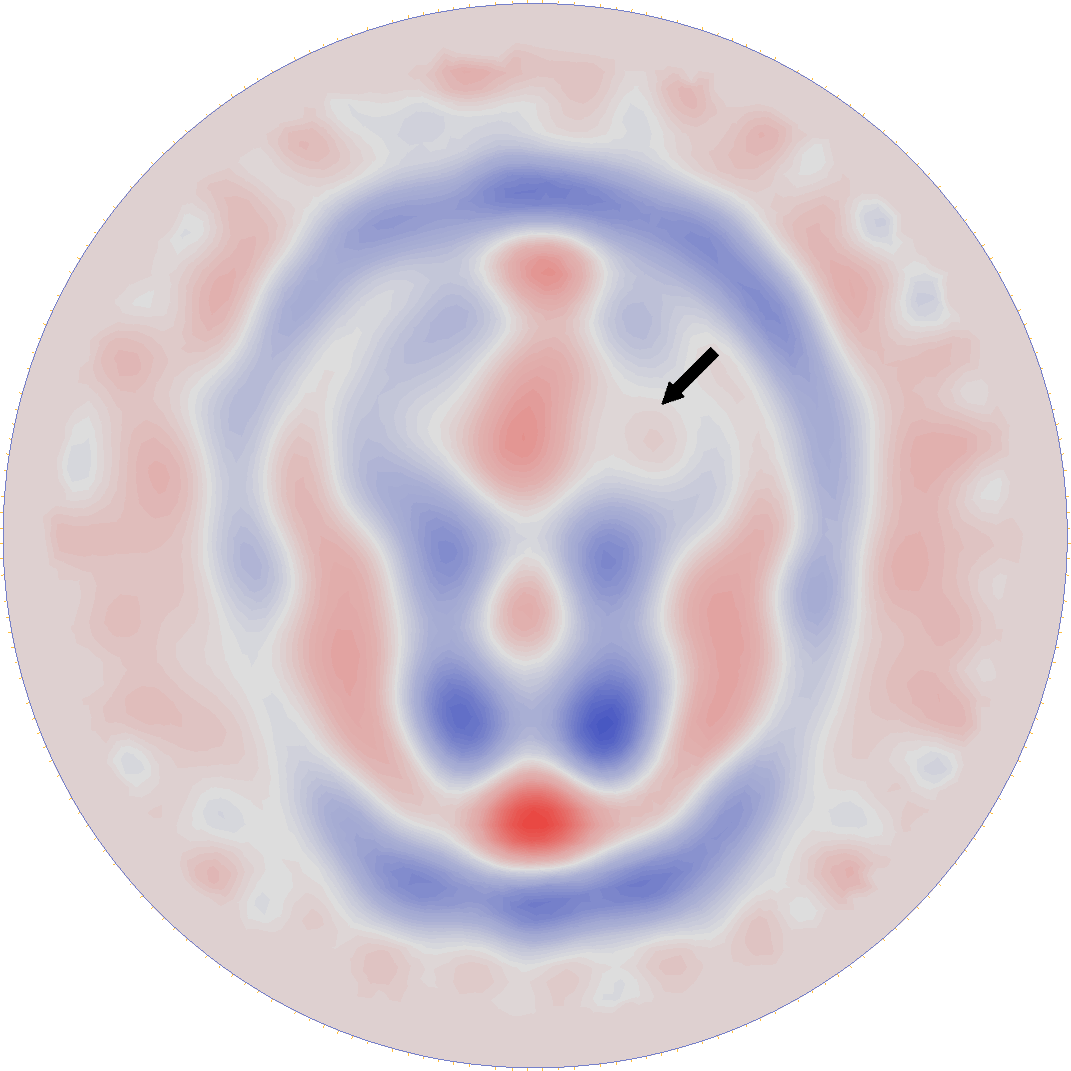}
\includegraphics[width = .3\textwidth]{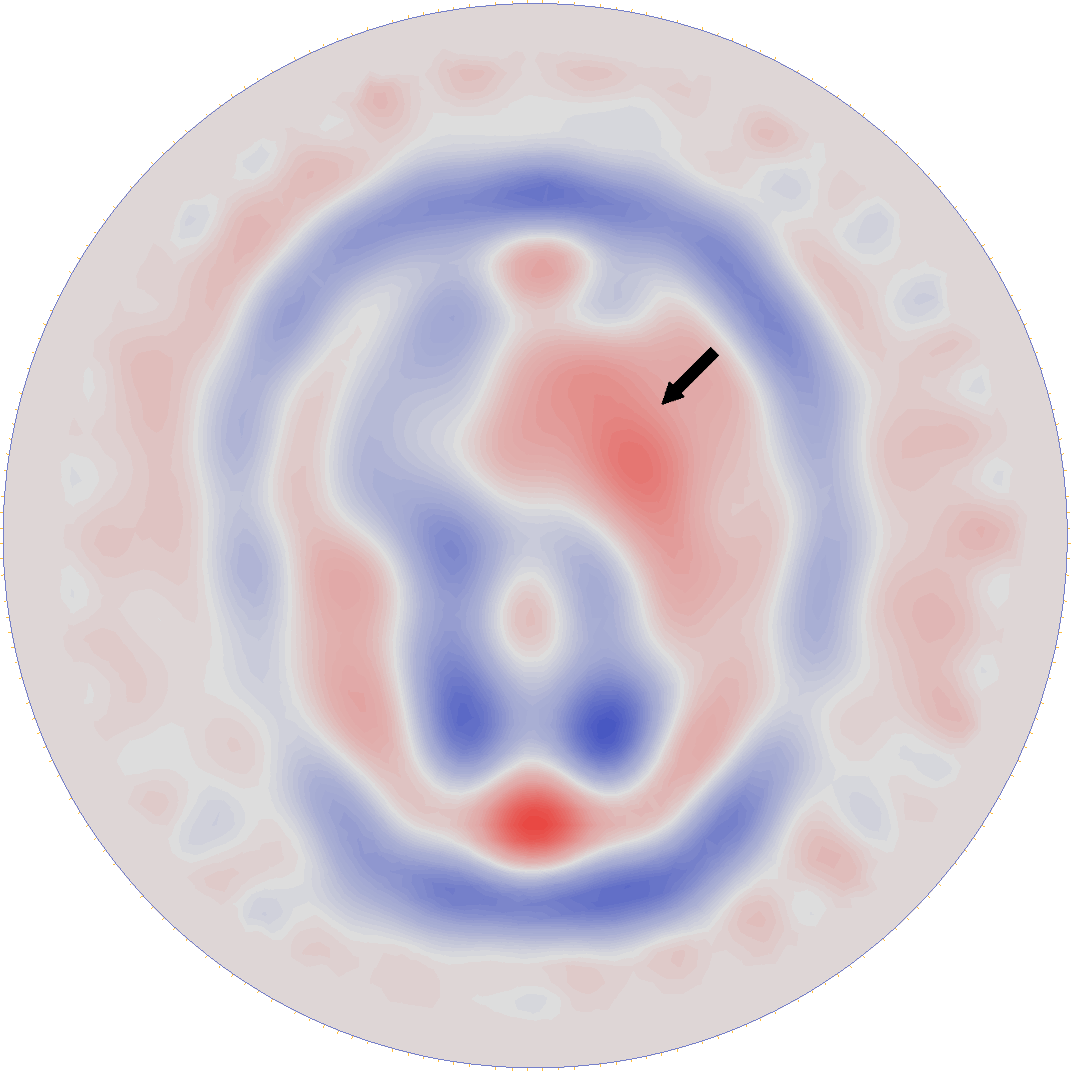}
%\caption{Evolution d'un AVC: permittivité exacte 3D et reconstruction avec des données bruitées (10\%)}
\caption{Top row: imaginary part of the exact permittivity for the healthy brain, small and large hemorrhagic strokes (indicated by the black arrow). Bottom row: corresponding reconstructions obtained by solving the inverse problem for the top layer.}
\label{fig:reconstruction}
\end{figure}

\paragraph{Reconstruction results for the top layer} We solve the inverse problem in the truncated domain containing only the first two rings of antennas from the top, and where only the coefficients $S_{ij}$ corresponding to transmitting antennas $j$ in the first ring are taken into account. The mesh of the computational domain is composed of \num{674580} tetrahedra, corresponding to approximately $10$ points per wavelength. The mesh size is twice as large as the one used for generating the synthetic data. The reconstruction process is faster when using a coarser mesh, and our numerical experiments have shown that using a finer mesh in the inverse problem does not improve the reconstruction. Each reconstruction starts from an initial guess consisting of the homogeneous matching solution and is obtained after reaching a convergence criterion of $10^{-2}$ for the value of the cost functional, which takes around $30$ iterations of the L-BFGS algorithm.

Figure~\ref{fig:reconstruction} shows the imaginary part of the exact and reconstructed permittivity for the three steps of the evolution of the hemorrhagic stroke, from the healthy brain (left column) to the large stroke (right column). Although it is well known that microwave imaging is not precise enough to resolve the very fine heterogeneities of the brain, we can see that the reconstructed images enable to track the evolution of the hemorrhagic stroke. More precisely, we can identify the appearance of the small stroke, even though the variations on the transmission coefficients between healthy brain and small stroke are very small as seen in Figure~\ref{fig:exactSij} (top).
It is difficult to assess quantitatively the quality of the reconstruction for low resolution imaging techniques such as microwave imaging; pointwise comparisons are not really meaningful. For this reason, we report in Table~\ref{tabperm} the mean value of the reconstructed  permittivity (exact permittivity in parentheses) in the ellipsoidal stroke region and its variation between healthy brain, small stroke and large stroke. We can see that although we do not quantitatively recover the exact values of the permittivity, the trend and order of magnitude of the variations are preserved in the reconstructions.

\begin{table}[h!]
\begin{center}
\begin{tabular}{|c|c|c|c|c|c|c|}
\hline 
\multirow{2}{*}{permittivity} & \multicolumn{2}{c|}{healthy brain}& \multicolumn{2}{c|}{small stroke} &  \multicolumn{2}{c|}{large stroke}\\
\cline{2-7}
 & real part & imag. part & real part & imag. part  & real part & imag. part \\
\hline
\multicolumn{1}{|c|}{mean value} & $43.2$ ($44.4$) & $15.4$ ($16.3$) & $45.7$ ($56.2$) & $18.6$ ($30.2$) & $51.6$ ($56.3$) & $23.6$ ($29.6$)\\
\hline
\multicolumn{1}{c|}{} &\multicolumn{2}{c|}{\multirow{2}{*}{variation}} & \multicolumn{2}{c|}{healthy $\rightarrow$ small} &  \multicolumn{2}{c|}{small $\rightarrow$ large}\\
\cline{4-7}
\multicolumn{1}{c|}{} &\multicolumn{2}{c|}{} & $+6\%$ ($+27\%$) & $+20\%$ ($+85\%$) & $+13\%$ ($+22\%$) & $+41\%$ ($+75\%$)\\
%\multicolumn{1}{|c||}{\% increase} & Re & Im & Re & Im\\
\cline{2-7}
\end{tabular} 
\end{center}
\caption{Mean value of the reconstructed permittivity in the ellipsoidal stroke region and variation of the permittivity between healthy brain and small stroke (in the small ellipsoid) and between small and large stroke (in the large ellipsoid). Values for the exact permittivity are reported in parentheses.}
\label{tabperm}
\end{table}

%Increasing the size of the ellipsoid in which the value of the permittivity is raised simulates the evolution of the stroke.

%$30$ iterations of the nonlinear conjugate gradient algorithm.
%As explained in Section~\ref{sec:inverseproblem}, we impose a first order absorbing boundary condition of Silver--M\"uller on the artificial boundary at the bottom of the truncated computational domain.

\begin{figure}[h!]
\centering 
\includegraphics[width = .48\textwidth]{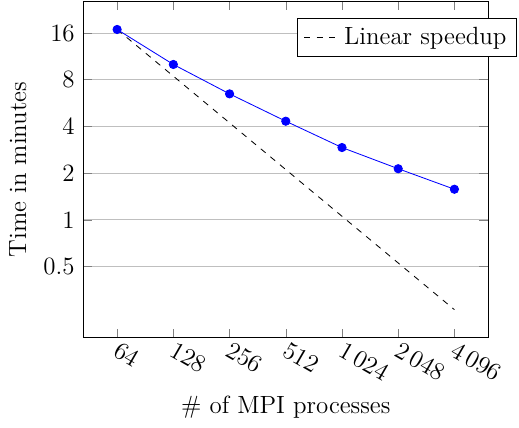}
\includegraphics[width = .48\textwidth]{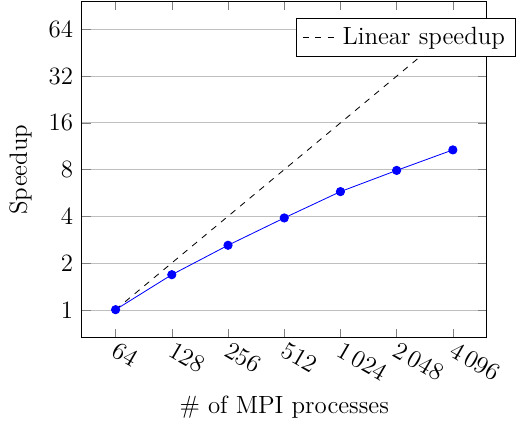}
%\caption{Temps de reconstruction en fonction du nombre de coeurs}
\caption{Strong scaling experiment: total time needed to obtain the third reconstructed image shown in Figure~\ref{fig:reconstruction} \textcolor{corrections}{(left) and corresponding speedup (right)}.}
\label{fig:scalinginverse}
\end{figure}
%Here we map one subdomain per MPI process, and use one thread per MPI process.
%to perform $30$ iterations of the conjugate gradient algorithm and obtain the reconstructed image

Figure~\ref{fig:scalinginverse} gathers the results of a strong scaling experiment which consists in solving the same inverse problem corresponding to the large stroke (third reconstructed image of Figure~\ref{fig:reconstruction}) for an increasing number of MPI processes. We report the total computing time needed to obtain the reconstructed image (left) and the corresponding speedup (right). We use one subdomain and one OpenMP thread per MPI process. To give an idea about the domain decomposition/pseudo-block tradeoff, we mention that the best computing time for $2048$ MPI processes is achieved by using $8$ domain decomposition communicators (i.e. $8$ concurrent direct solves) with $256$ subdomains treating $4$ right-hand sides each. In contrast to the strong scaling experiment for the direct problem in Section~\ref{sec:scaling}, we obtain sublinear speedups for the inverse problem. This can be explained by the fact that the solution phase dominates the overall cost, with the setup phase being less prevalent. Indeed, linear systems to be solved in the inverse problem are smaller, with multiple right-hand sides each. We thus observe deteriorating efficiency, as the one-level domain decomposition preconditioner is not perfectly numerically scalable (increase in terms of number of GMRES iterations) at large process counts, where the setup cost is the smallest.

Nevertheless, Figure~\ref{fig:scalinginverse} shows that we can generate an image with a total computing time of less than $2$ minutes ($94$ seconds) using $4096$ cores. These preliminary results are very encouraging as they show that we are able to achieve a satisfactory reconstruction time in the perspective of using such an imaging technique for monitoring. 
%Indeed, doctors would like to have an image every $15$ minutes. 
This allows clinicians to obtain almost instantaneous images 24/7 or on demand. Although the reconstructed images do not feature the complex heterogeneities of the brain, which is in accordance with what we expect from microwave imaging methods, they allow the characterization of the stroke and its monitoring, at least on synthetic data.

\color{black}

\section{Conclusion}
\label{sec:conclusion}

We have developed a tool that reconstructs a microwave tomographic image of the brain in less than $2$ minutes using $4096$ cores. This computational time corresponds to clinician acceptance for rapid diagnosis or medical monitoring at the hospital. These images were obtained from noisy synthetic data from a very accurate model of the brain. To our knowledge, this is the first time that such a realistic study (operational acquisition device, highly accurate three-dimensional synthetic data, 10\% noise) shows the feasibility of microwave imaging. This study was made possible by the use of massively parallel computers and facilitated by the HPDDM and FreeFem++ tools that we have developed. The next step is the validation of these results on clinical data.

Regarding the numerical aspects of this work, we will accelerate the solution of the series of direct problems, which accounts for more than 80\% of our elapsed time. We explain here the three main avenues of research:
\begin{itemize}
\item The present ORAS solver for Maxwell's equations is a one level algorithm, which cannot scale well over thousands of subdomains. The introduction of a two-level preconditioner with an adequate coarse space would allow for very good speedups even for decompositions into a large number of subdomains.
\item Recycling information obtained during the convergence of the optimization algorithm will also enable us to improve the performance of the method, see~\cite{Parks:2006:RKS}.
\item Iterative block methods that allow for simultaneous solutions of linear systems have not been fully investigated. Arithmetic intensity would be increased since block methods may converge in a smaller number of iterations while exploiting modern computer architectures effectively.
\end{itemize}


\section*{Acknowledgments}
This work was granted access to the HPC resources of TGCC@CEA under the allocations 2016-067519 and 2016-067730 made by GENCI. This work has been supported in part by ANR through project MEDIMAX, ANR-13-MONU-0012.

\bibliographystyle{elsarticle-num} 
\bibliography{paperPP_review}


%\begin{thebibliography}{00}
%%% Text of bibliographic item
%
%\bibitem{}
%
%\end{thebibliography}

\end{document}